\documentclass[12pt,a4paper]{article}

\usepackage{amsmath}
\usepackage{amssymb}
\usepackage[nosort]{cite}
\def\gfxon{\usepackage[final]{graphicx}}

\gfxon

\makeatletter
\let\old@makecaption=\@makecaption
\def\@makecaption{\small\old@makecaption}
\makeatother

\makeatletter
\@addtoreset{equation}{section}
\makeatother

\makeatletter
\let\old@startsection=\@startsection
\renewcommand{\@startsection}[6]{\old@startsection{#1}{#2}{#3}{#4}{#5}{#6\mathversion{bold}}}
\makeatother

\let\oldPhi=\Phi
\let\oldPsi=\Psi
\let\oldGamma=\Gamma
\let\oldSigma=\Sigma
\renewcommand{\Phi}{\mathnormal{\oldPhi}}
\renewcommand{\Psi}{\mathnormal{\oldPsi}}
\renewcommand{\Gamma}{\mathnormal{\oldGamma}}
\renewcommand{\Sigma}{\mathnormal{\oldSigma}}

\newcommand{\hypref}[2]{\ifx\href\asklfhas #2\else\href{#1}{#2}\fi}
\newcommand{\Secref}[1]{Section~\ref{#1}}
\newcommand{\secref}[1]{Sec.~\ref{#1}}

\newcommand{\appref}[1]{App.~\ref{#1}}

\newcommand{\tabref}[1]{Tab.~\ref{#1}}

\newcommand{\figref}[1]{Fig.~\ref{#1}}

\newcommand{\sfrac}[2]{{\textstyle\frac{#1}{#2}}}
\newcommand{\half}{\sfrac{1}{2}}
\newcommand{\quarter}{\sfrac{1}{4}}

\newcommand{\grSU}{\mathrm{SU}}

\newcommand{\grU}{\mathrm{U}}
\newcommand{\grSO}{\mathrm{SO}}
\newcommand{\grSp}{\mathrm{Sp}}
\newcommand{\order}[1]{\mathcal{O}(#1)}

\newcommand{\superN}{\mathcal{N}}
\newcommand{\gym}{g_{\scriptscriptstyle\mathrm{YM}}}
\newcommand{\gnorm}{g}

\newcommand{\Tr}{\mathop{\mathrm{Tr}}}

\newcommand{\rep}[1]{{\mathbf{#1}}}

\newcommand{\indup}[1]{_{\mathrm{#1}}}

\newcommand{\tsum}{{\textstyle\sum}}

\newcommand{\trans}{{\scriptscriptstyle\mathsf{T}}}
\newcommand{\conj}{\ast}

\newcommand{\Integers}{\mathbb{Z}}

\newcommand{\cder}{\mathcal{D}}
\newcommand{\matr}[2]{\left(\begin{array}{#1}#2\end{array}\right)}
\newcommand{\Op}{\mathcal{O}}

\newcommand{\OpV}{\mathcal{Q}}

\newcommand{\lrbrk}[1]{\left(#1\right)}
\newcommand{\bigbrk}[1]{\bigl(#1\bigr)}
\newcommand{\brk}[1]{(#1)}

\newcommand{\normord}[1]{\mathopen{:}#1\mathclose{:}}

\newcommand{\bigvev}[1]{\bigl\langle#1\bigr\rangle}
\newcommand{\bigcomm}[2]{\big[#1,#2\big]}
\newcommand{\comm}[2]{[#1,#2]}

\newcommand{\acomm}[2]{\{#1,#2\}}

\newcommand{\bigabs}[1]{\bigl|#1\bigr|}
\newcommand{\bigeval}[1]{#1\big|}
\newcommand{\eval}[1]{#1|}

\newcommand{\nn}{\nonumber}
\newcommand{\nln}{\nonumber\\}
\newcommand{\nl}{\nonumber\\&&\mathord{}}
\newcommand{\nlnum}{\\&&\mathord{}}
\newcommand{\nle}{\nonumber\\&=&\mathrel{}}
\newcommand{\eq}{\mathrel{}&=&\mathrel{}}
\newenvironment{myeqnarray}{\arraycolsep0pt\begin{eqnarray}}{\end{eqnarray}\ignorespacesafterend}
\newenvironment{myeqnarray*}{\arraycolsep0pt\begin{eqnarray*}}{\end{eqnarray*}\ignorespacesafterend}

\def\[{\begin{equation}}
\def\]{\end{equation}}
\def\<{\begin{myeqnarray}}
\def\>{\end{myeqnarray}}

\ifx\href\asklfhas\newcommand{\href}[2]{#2}\fi
\newcommand{\arxivno}[1]{\href{http://arxiv.org/abs/#1}{#1}}

\begin{document}

\thispagestyle{empty}
\begin{flushright}\footnotesize
\texttt{\arxivno{hep-th/0303060}}\\
\texttt{AEI 2003-028}
\end{flushright}
\vspace{2cm}

\begin{center}
{\Large\textbf{\mathversion{bold}The Dilatation Operator of Conformal
\\$\mathcal{N}=4$ Super Yang-Mills Theory}\par}
\vspace{2cm}

\textsc{N. Beisert, C. Kristjansen and M. Staudacher}
\vspace{5mm}

\textit{Max-Planck-Institut f\"ur Gravitationsphysik\\
Albert-Einstein-Institut\\
Am M\"uhlenberg 1, D-14476 Golm, Germany}
\vspace{3mm}

\texttt{nbeisert,kristjan,matthias@aei.mpg.de}\par\vspace{2cm}

\textbf{Abstract}\vspace{7mm}

\begin{minipage}{14.7cm}
We argue that existing methods for the perturbative 
computation of anomalous dimensions and the disentanglement of
mixing in ${\cal N}=4$ gauge theory
can be considerably simplified, systematized and extended by 
focusing on the theory's dilatation operator. 
The efficiency of the
method is first illustrated at the one-loop level for general 
non-derivative scalar states. 
We then go on to derive, for pure scalar states, the 
two-loop structure of the dilatation operator. This allows us to obtain
a host of new results. Among these are an infinite number of
previously unknown two-loop anomalous dimensions,  
new subtleties concerning 't Hooft's large $N$ expansion due 
to mixing effects of degenerate single and multiple trace states, 
two-loop tests of various protected operators,
as well as two-loop non-planar results for two-impurity operators 
in BMN gauge theory.
We also put to use the recently discovered integrable spin chain
description of the planar one-loop dilatation operator
and show that the associated
Yang-Baxter equation explains the existence of a hitherto unknown
planar ``axial'' symmetry between infinitely many gauge theory states.
We present evidence that this integrability can be extended to
all loops, with intriguing consequences for gauge theory,
and that it leads to a novel integrable deformation of 
the XXX Heisenberg spin chain.
Assuming that the integrability
structure extends to more than two loops, we determine 
the planar three-loop contribution to the dilatation operator.
\end{minipage}

\end{center}

\newpage
\setcounter{page}{1}
\setcounter{footnote}{0}

\section{Introduction and overview}
\label{sec:intro}

Conformal quantum field theories have been fascinating theoretical
physicists for a long time. In two dimensions they are arguably
the most important class of field theories: Mathematically, 
they are quite tractable once one puts to full use the fact that
the conformal algebra has an infinite number of
generators. In consequence many exact results on the spectrum
of operators and the structure of correlation functions may be
derived. Apart form their pivotal importance to string theory,
they are physically of great value due to their relationship
with critical phenomena and integrable models of statistical
mechanics. For example, in many cases the representation theory
of 2D conformal field theories fixes the scaling dimensions of
local operators, which in turn are often related to critical exponents
of experimentally relevant systems of solid states physics.

In four dimensions conformal symmetry was long believed to play only a
minor role. The group has only finitely many generators, the
QFT's relevant to particle physics are certainly not conformal,
and the only, trivial, example seemed to be free, massless field theory.
However, after the discovery of supersymmetry, it has become clear
that supersymmetric gauge theories can be exactly conformally
invariant on the quantum level, and in many cases the phase diagram
of such gauge theories contained conformal points or regions,
quite analogously to phase diagrams in two dimensions. There is 
therefore obvious theoretical interest in increasing our understanding
of these phases. In particular, the 4D gauge theory with the maximum
possible number $\superN=4$ of rigid supersymmetries, discovered in 1976 
\cite{Gliozzi:1977qd,Brink:1977bc}, has a superconformal phase
\cite{Sohnius:1981sn,Howe:1984sr,Brink:1983pd}.

Supersymmetric gauge theories are intimately connected to
superstring theory. In fact, the $\superN=4$ action was originally
discovered \cite{Gliozzi:1977qd} by considering the low-energy 
limit of superstrings. Surprisingly, with the advent of the AdS/CFT
correspondence (see \cite{Aharony:1999ti,D'Hoker:2002aw} 
for comprehensive reviews) it 
was argued that the $\superN=4$ gauge theory \emph{is}, via duality, 
a superstring theory in a particular background. This conjecture
woke the Sleeping Beauty and resulted in a large number of 
investigations into the structure of $\superN=4$. 
In particular the representation theory of the superconformal symmetry
group $\grSU(2,2|4)$ \cite{Dobrev:1985qv} was investigated more closely  
\cite{Andrianopoli:1998ut,Andrianopoli:1999vr}, and numerous
non-renormalization theorems were derived, see e.g.~\cite{Lee:1998bx}.
In addition, some unexpected non-renormalization theorems 
that do not follow from $\grSU(2,2|4)$ representation theory
were found \cite{Arutyunov:2000ku}. Once thought to be somewhat boring,
it gradually became clear that conformal $\superN=4$ is an extremely rich
and non-trivial theory with many hidden secrets. We will see more
examples for its intricate structure in the present paper.

While many of the recently obtained new results on $\superN=4$
in principle could have been derived a long time ago, 
the AdS/CFT correspondence clearly helped immensely in 
formulating the right questions. The latest chapter in exploiting
this useful duality involved taking a certain limit on both sides
of the correspondence: On the string side, after a process termed
Penrose contraction of the AdS space, one obtains a new maximally
supersymmetric background for the IIB 
superstring~\cite{Blau:2001ne,Blau:2002dy}
which has the great advantage that the spectrum of free, massive
string modes can be found exactly
\cite{Metsaev:2001bj,Metsaev:2002re}.
On the gauge theory side, one considers the so-called BMN limit 
\cite{Berenstein:2002jq}.
It involves studying infinite sequences of operators 
containing an increasing number of fields.
The relationship to plane wave string excitations is then
made by comparing the conformal scaling dimension of these
high-dimension operators to the masses of string excitations
\cite{Berenstein:2002jq}. One therefore had to develop techniques
to efficiently compute in perturbation theory
the scaling dimensions of operators containing
an arbitrary number of fields. This was done in various papers,
on the planar \cite{Berenstein:2002jq,Gross:2002su,Parnachev:2002kk,
Beisert:2002tn,Klose:2003tw} and non-planar
level \cite{Kristjansen:2002bb,Constable:2002hw,
Gursoy:2002yy, Beisert:2002bb,Constable:2002vq,
Eynard:2002df,Gursoy:2002fj},
extending earlier work on protected half BPS 
\cite{D'Hoker:1998tz,Penati:1999ba,Penati:2000zv,Penati:2001sv}
and quarter BPS operators \cite{Ryzhov:2001bp}.
In \cite{Beisert:2002ff} it was realized, following important insights by
\cite{Gross:2002mh,Janik:2002bd} (see also \cite{Minahan:2002ve}) 
that these well-established
techniques can be considerably simplified and extended, 
as we will now explain.

The standard way 
to find the scaling dimensions $\Delta_\alpha$ 
of a set of conformal fields
$\hat\Op_\alpha$, employed by all of the just mentioned papers,
is to consider the two-point functions
\[
\bigvev{\hat\Op_\alpha(x)\,  \hat\Op_\beta(0)}= 
\frac{\delta_{\alpha \beta}}{|x|^{ 2\Delta_\alpha}}.
\label{two}
\]
Here the form of the two-point function is determined by conformal
symmetry alone. Finding the scaling dimension $\Delta_\alpha$ 
of a conformal field 
$\Op_\alpha$ is more subtle and, in general, requires an understanding
of the dynamics of the theory. In the free field limit 
(i.e.~zero coupling constant $\gym =0$) $\Delta$ equals the naive
classical (tree level) scaling dimension obtained by standard power counting.
For weak coupling $\gym$, we can compute corrections to the naive dimension
by perturbation theory.  
In practice, extracting these corrections 
from eq.\eqref{two} is somewhat tedious and fraught with various
technical complications. For one, while the corrections are
perfectly \emph{finite} numbers, the perturbative computation of
two-point functions such as eq.\eqref{two}
leads to spurious infinities, requiring renormalization%
\footnote{An alternative method consists in extracting
the anomalous dimensions from four-point functions. This is
how e.g.~the two-loop anomalous dimension of the Konishi field
was first found \cite{Bianchi:2000hn,Arutyunov:2001mh}.
However, this is not really different, as one
can look at these four-point functions,
via a double operator product expansion,
 as a generator of 
two-point functions regulated by a 
different method, namely point-splitting.}.
Secondly, the definition of the scaling dimensions tacitly assumes
that we have already found the correct \emph{conformal} operators
$\hat\Op_\alpha$. However, if one starts with a set of naive operators
$\Op_\alpha$ with the same engineering (tree-level) dimension 
one generically encounters the phenomenon of mixing:
The two-point function is not diagonal in $\alpha,\beta$, 
and one rather has
a matrix $\langle \Op_\alpha(x) \Op_\beta(0) \rangle$ 
since a generic field does not have a definite scaling dimension.
It therefore seems that we have to diagonalize
the two-point functions, after renormalization,
order by order in perturbation theory.

Here we would like to shift the attention away from the 
two-point functions eq.\eqref{two} and rather focus on the
dilatation operator acting on states at the origin of space-time
(in a radial quantization scheme) as was already done, for one-loop, 
in \cite{Minahan:2002ve} (on the planar level) and
in \cite{Beisert:2002ff} (in the BMN limit).
Its eigensystem consists of the 
eigenvalues $\Delta_\alpha$ and the eigenstates $\hat\Op_\alpha$.
One thus has
\[
D\, \hat\Op_\alpha=\Delta_\alpha\,\hat\Op_\alpha.
\]
By way of example we will restrict the discussion to fields
solely composed of the six scalar fields 
$\Phi_n=\Phi_n^{(a)}T^a$ ($n=1,\ldots,6$,
$a=1,\ldots,N^2-1$)
of the $\superN=4$ $\grSU(N)$ gauge theory, 
but the extension to a larger class
of fields is clearly possible. (Our notation and conventions
are explained in \appref{sec:conv}.)
The classical dilatation operator $D_0$ 
then simply counts the total number of scalar fields,
as the engineering dimension of scalars is one.
This can be formally written as 
\[\label{eq:D0}
D_0=\Tr \Phi_m \check\Phi_m,
\]
where we introduced the notation $\check \Phi$
for the variation with respect to the field $\Phi$
\footnote{In the language of canonical 
quantization the field $\Phi$ and the variation $\check\Phi$ correspond 
to creation and annihilation operators, respectively.
The dilatation generator corresponds to the Hamiltonian
and a vacuum expectation value is obtained by
setting the fields to zero, $\eval{\ldots}_{\Phi=0}$.}
\[
\label{check}
\check \Phi_m=\frac{\delta}{\delta \Phi_m}=
T^a \frac{\delta}{\delta \Phi^{(a)}_m}.
\]
In an interacting theory the scaling dimensions
and therefore also the dilatation generator depend on 
the coupling constant. 
In perturbation theory the dilatation generator can be 
expanded in powers of the coupling constant
\[
D=\sum_{k=0}^{\infty} \left( \frac{\gym^2}{16\pi^2} \right)^k D_{2 k},
\label{dilexp}
\]
and we shall denote $D_{2k}$ as the $k$-loop dilatation generator.
The full one-loop contribution $D_2$ was first worked out, in the
disguise of an ``effective vertex'', in appendix C of
\cite{Beisert:2002bb} (but see also 
\cite{Kristjansen:2002bb,Constable:2002hw,Constable:2002vq,Beisert:2002tn,
Minahan:2002ve,Beisert:2002ff}):
\footnote{Note that this expression is 
perfectly valid also for $\grSO(N)$, $\grSp(N)$ and exceptional
gauge groups.}
\[\label{eq:D2}
D_2= 
-\normord{\Tr \comm{\Phi_m}{\Phi_n}\comm{\check\Phi_m}{\check\Phi_n}}
-\half\normord{\Tr \comm{\Phi_m}{\check\Phi_n}\comm{\Phi_m}{\check\Phi_n}}.
\]
However, it was shown in \cite{Beisert:2002ff} that one 
can \emph{bypass} the consideration of two-point functions and directly
diagonalize $D_2$ in any convenient, linearly independent and complete basis.
In eq.\eqref{eq:D2} the normal ordering notation indicates that
the derivatives do not act on the fields enclosed by $\normord{\phantom{x}}$. 
To see how $D_2$ acts, let us consider the simplest scalar fields
of the theory, which have classical dimension two. 
One has the protected chiral primaries $\OpV_{nm}$
as well as the Konishi field $\mathcal{K}$
\[
\label{dimtwo}
\OpV_{nm}=\Tr \Phi_n \Phi_m -\sfrac{1}{6} \delta_{nm} \Tr \Phi_k \Phi_k
\qquad {\rm and} \qquad {\cal K}=\Tr\Phi_k \Phi_k.
\]
One immediately verifies%
\footnote{The fission rule 
$\Tr A \check \Phi_m B \Phi_n = \delta_{mn}\Tr A \Tr B$ and the fusion rule
$\Tr A \check \Phi_m \Tr \Phi_n B =\delta_{mn}\Tr A B$
are useful when calculating the action of $D_2$.
These are the $\grU(N)$ rules but due to the commutators in 
$D_2$ there is no difference when using the $\grSU(N)$ rules.}, 
using 
eqs.\eqref{eq:D0},\eqref{check},\eqref{dilexp},\eqref{eq:D2}
that, to one loop, as $D_2 \OpV_{nm}=0$ and $D_2 \mathcal{K}=12 N\, \mathcal{K}$,
\[
\label{konishi}
\Delta_{\OpV}=2 \qquad \mbox{and} \qquad 
\Delta_{\mathcal{K}}=2+\frac{3 \gym^2 N}{4 \pi^2},
\]
that is the one-loop anomalous dimension of $\OpV_{nm}$
vanishes as expected, while the anomalous dimension of the Konishi
scalar agrees with the well-known result \cite{Anselmi:1997mq}.

Let us next look at a less trivial example and consider the following
set of unprotected $\grSO(6)$ invariant operators of classical
dimension four, spanning a four-dimensional space of states: 
\[
\label{glebs}
\Op=\left(\begin{array}{c}
\Tr \Phi_m\Phi_n \Tr\Phi_m\Phi_n\\
\Tr \Phi_m\Phi_m \Tr\Phi_n\Phi_n\\\hline
\Tr \Phi_m\Phi_n \Phi_m\Phi_n\\
\Tr \Phi_m\Phi_m \Phi_n\Phi_n
\end{array}\right).
\]
For the convenience of the reader we are optically separating 
double and single trace sectors by solid lines. 
This case was studied in detail in \cite{Arutyunov:2002rs}, using
the standard procedure (see also \cite{Penati:2001sv})
of directly analyzing, employing ${\cal N} =1$ superspace Feynman rules, 
the set of two-point functions of the four composite fields,
\emph{cf}.\ our above discussion surrounding eq.\eqref{two}.  
Using eq.\eqref{eq:D2}, we find the following action of the 
one-loop dilatation generator
\[
\label{glebsmat}
D_2=N\left(\begin{array}{cc|cc}
0             &            4&-\sfrac{20}{N}&\sfrac{20}{N}\\
0             &           24&-\sfrac{24}{N}&\sfrac{24}{N}\\\hline
-\sfrac{24}{N}& \sfrac{4}{N}&             8&-4\\
- \sfrac{4}{N}&\sfrac{14}{N}&            -4&18
\end{array}\right).
\]
The matrix notation indicates the fields that are generated by the action
of $D_2$ from the ones in eq.\eqref{glebs}, e.g.
\<
D_2 \Tr \Phi_m\Phi_n \Tr\Phi_m\Phi_n \eq
  0 \Tr \Phi_m\Phi_n \Tr\Phi_m\Phi_n
+4 N \Tr \Phi_m\Phi_m \Tr\Phi_n\Phi_n 
\nl
-20 \Tr \Phi_m\Phi_n \Phi_m\Phi_n
+20 \Tr \Phi_m\Phi_m \Phi_n\Phi_n.
\>
We see that the states in eq.\eqref{glebs} are \emph{not} eigenstates
of the dilatation operator: The latter linearly mixes
all operators in the four-dimensional vector space\cite{Arutyunov:2002rs}.
However, diagonalizing the dilatation
matrix $D_2$ (we will throughout this paper take the freedom of using 
the same notation $D_{2 k}$ for both the dilatation operator and
its matrix representations) in eq.\eqref{glebsmat} will in one go
resolve the mixing problem and yield the anomalous dimensions:
The conformal operators are given, up to normalization, by the
eigenvectors of eq.\eqref{glebsmat} (in the basis eq.\eqref{glebs}),
and their respective one-loop anomalous dimensions correspond to the 
eigenvalues of eq.\eqref{glebsmat}. The latter are obtained by
finding the roots of the characteristic polynomial
$\det ( \omega - \half D_2 )$:
\[
\omega^4-25\omega^3+\lrbrk{188-\frac{160}{N^2}}\omega^2
-\lrbrk{384-\frac{1760}{N^2}}\omega-\frac{7680}{N^2}=0.
\]
The dimensions of the four conformal operators are
\[\Delta=4+\frac{\gym^2N}{8\pi^2}\,\omega,\]
with $\omega$ being the four different roots of the above quartic
equation. 
%
They are easily seen to agree with the ones obtained in 
\cite{Arutyunov:2002rs}. Comparison with the methodology of
\cite{Arutyunov:2002rs} clearly demonstrates the 
great simplifications resulting from the direct use of
the dilatation operator%
\footnote{Incidentally, one easily verifies that our
dilatation matrix $D_2$ in eq.\eqref{glebsmat} is, up to an overall
constant, precisely given by the product of the one-loop mixing
matrix eq.(5.4) of \cite{Arutyunov:2002rs}
times the inverse of the tree-level mixing matrix
eq.(5.3) of \cite{Arutyunov:2002rs}, in agreement with the results of
\cite{Beisert:2002ff}, where the relation between the
traditional method and the dilatation operator method was explained
in detail. This furnishes a nice, explicit example illustrating
that much of the 
combinatorial complications of the usual method cancel out from
the physics.}.

In this paper we will find the two-loop contribution $D_4$ to
the scalar dilatation operator $D$ eq.\eqref{dilexp}, with the
restriction to operators in the $\grSO(6)$ 
representations $[p,q,p]$ with tree-level scaling dimension $2p+q$.
These fields may always be constructed out of two complex scalars, say
$ Z=\sfrac{1}{\sqrt{2}}\brk{\Phi_5+i\Phi_6}$ and
$\phi=\sfrac{1}{\sqrt{2}}\brk{\Phi_1+i\Phi_2}$.
We will argue that operators composed by these two fields
form a closed set under dilatation. 
The idea is then to inspect
all Feynman diagrams contributing to the dilatation operator
and write down the possible linear combinations of terms
contributing to $D_4$. We fix some unknown constants by using
gauge invariance, and proven two-loop non-renormalization theorems
\cite{Penati:1999ba,Penati:2000zv}.
To save us extra work, we finally use
a result conjectured in the context of BMN gauge theory in 
\cite{Berenstein:2002sa} and derived in 
\cite{Gross:2002su,Santambrogio:2002sb}
to determine the remaining coefficients.
All this is done without explicitly working out a single Feynman diagram!
One finds
\<\label{introD4}
D_0\eq
\Tr Z\check Z+\Tr \phi\check\phi,
\nln
D_2\eq
-2\,\normord{\Tr \comm{\phi}{Z}\comm{\check\phi}{\check Z}},
\nln
D_4\eq
-2\, \normord{\Tr \bigcomm{\comm{\phi}{Z}}{\check Z}\bigcomm{\comm{\check\phi}{\check Z}}{Z}}
\nl
-2\, \normord{\Tr \bigcomm{\comm{\phi}{Z}}{\check \phi}\bigcomm{\comm{\check\phi}{\check Z}}{\phi}}
\nl
-2\, \normord{\Tr \bigcomm{\comm{\phi}{Z}}{T^a}\bigcomm{\comm{\check\phi}{\check Z}}{T^a}},
\>
where we have also explicitly rewritten the action of $D_0$ and $D_2$ 
on two complex scalars. 
Let us check eq.\eqref{introD4} and apply our result to work out the two-loop
dimension of the Konishi field. We cannot directly apply $D_4$ 
to ${\cal K}$ in eq.\eqref{dimtwo} as we have not yet worked out the
terms relevant to scalar fields containing $\grSO(6)$ traces.
However, it is well-known that the following field is a
Konishi descendant, with identical anomalous correction to its
classical dimension four:
\[
\mathcal{K}'=
\Tr \comm{\phi}{Z}\comm{\phi}{Z}.
\]
Applying eq.\eqref{introD4} to ${\cal K}'$ we find
\[\label{Kontwoloop}
\Delta_{\mathcal{K}'} = 4+\frac{3 \gym^2 N}{4\pi^2}
-\frac{3 \gym^4 N^2}{16\pi^4}.
\]
This agrees with the results of 
\cite{Bianchi:2000hn,Arutyunov:2001mh}.

Recently it was discovered by Minahan and Zarembo
that properties of pure scalar states in
the planar, one-loop sector of $\superN=4$
Super Yang-Mills theory can be described by an
\emph{integrable} $\grSO(6)$ spin chain \cite{Minahan:2002ve}:
The planar version of the one-loop dilatation operator
maps onto the spin chain Hamiltonian. 
Interestingly, we are able to derive from this spin
chain picture new results for the gauge theory. 
The integrability ensures, via the existence of an $R$-matrix
satisfying the Yang-Baxter equation,
the existence of further conserved charges in addition to the Hamiltonian. 
Here we shall find
that the first non-trivial conserved charge of the spin chain translates
into an ``axial'' symmetry of $\superN=4$ Super Yang-Mills theory
at $N=\infty$. 
The latter transforms between states of opposite ``parity'' 
with respect to reflection of color traces.

What is more, this symmetry turns
out to be conserved by the two loop interactions as well.
This might indicate that ${\cal N}=4$ gauge theory can be used
as a source for new integrable spin chain Hamiltonians with 
interactions more subtle than just the standard nearest neighbor ones:
At one-loop, we have the standard XXX$_{1/2}$ Heisenberg spin chain,
which is exactly integrable.
Higher loop quantum corrections constitute deformations of the spin chain
with a smooth deformation parameter $\gym^2$. 
At $\order{\gym^{2k}}$ 
the spin chain interactions involve $k$ nearest neighbors.
When all interactions up to $\order{\gym^{2k}}$ are included,
the model's charges commute only up to terms of higher order.
Therefore, in order for the deformed model to still 
be \emph{exactly} integrable,
\emph{all} quantum corrections (any $k$) have to be taken into account. 
Our integrable deformation is thus, 
for finite $\gym$, non-local.
In a certain sense locality is recovered as $\gym$ is tuned to zero.

Assuming the symmetry to hold
at the three-loop level allows us to fix the planar version of
the three-loop dilatation operator for scalar operators of the type
discussed above. E.g.~the resulting dilatation operator implies the following
result for the three-loop planar contribution to $\Delta_{{\cal K}'}$:
\[\label{Konthreeloop}
\Delta_{{\cal K}'}
=4+\frac{3 \gym^2 N}{4\pi^2}
-\frac{3 \gym^4 N^2}{16\pi^4}
+\frac{21\gym^6 N^3}{256\pi^6}.
\]

Our paper is organized as follows. We start by recalling the derivation
of the one-loop dilatation generator in \secref{sec:oneloop} and move on
to determine its two-loop version in
\secref{sec:twoloop}. \Secref{sec:lowdim} is devoted to the
application of the resulting dilatation generator to lower dimensional
operators and contains in particular a discussion of certain
peculiarities of the large-$N$ expansion related to degeneracy and
mixing of single and multiple trace states. Next follows in 
\secref{sec:twoimp} a treatment of two-impurity operators of
arbitrary finite dimension in the spirit of
reference~\cite{Beisert:2002tn}, leading in particular to an exact
expression for the planar two-loop anomalous dimension for any such operator.
As a corollary we show in \secref{sec:BMNlimit} that the two-loop
contribution to the quantum mechanical Hamiltonian describing BMN
gauge theory~\cite{Beisert:2002ff}
 can be simply expressed in terms of its one-loop
counterpart which suggests an all genera version of the celebrated
all-loop BMN square root 
formula~\cite{Berenstein:2002jq,Gross:2002su,Santambrogio:2002sb}. 
In \secref{sec:NoMystery} we discuss the 
spin chain description of
${\cal N}=4$ Super Yang-Mills theory~\cite{Minahan:2002ve} and point
out certain consequences thereof, \emph{cf}.\ the discussion above.
Finally, in \secref{sec:highergrounds} we
discuss possible implications for
arbitrarily high loop orders. 
In particular, we speculate about how fully 
exploiting the integrability might lead to exact planar anomalous
dimensions. In addition we conjecture the existence of 
novel integrable deformations of the standard Heisenberg spin chain.
We furthermore call the attention of the reader to
\appref{sec:anodim} which contains a collection of previously
unknown anomalous dimensions for lower dimensional operators.

\section{One-loop revisited}
\label{sec:oneloop}

The generator of dilatations on scalar fields at one-loop is given by
eq.\eqref{eq:D2}.
At one-loop mixing between operators with different kinds of fields
is irrelevant. Therefore \eqref{eq:D2} alone determines
the one-loop anomalous dimensions of 
operators whose leading piece is purely made out of scalar fields.

In the following we will rederive this result to make the reader 
familiar with the methods and transformations in a rather simple context.
All these steps will reappear at two-loops in a more involved fashion.
We start by introducing the notation at tree-level. We sketch
the renormalization procedure at one-loop and apply it to
obtain the one-loop dilatation generator.

\subsection{Tree-level}
\label{ssec:treelevel}

We obtain the anomalous dimensions of a set of operators $\Op_\alpha$ 
by formally evaluating their two-point functions.
For that we distinguish between fields at points
$x$ and $0$ by the superscript $\pm$
\[\label{eq:+-Notation}
\Phi_m^+=\Phi_m(x),\quad\Phi_m^-=\Phi_m(0).\]
The operators $\Op_\alpha$ are constructed as traces of scalar fields
\[\Op_\alpha(\Phi)=\Tr \Phi_m\Phi_n\Phi_m\ldots \,\Tr\Phi_n\ldots,
\qquad \Op^\pm_\alpha=\Op_\alpha(\Phi^\pm),\]
where $\alpha$ enumerates the operators.
The tree-level two-point function can formally be written as
\[\label{eq:treelevelW}
\bigvev{\Op_\alpha^+ \Op_\beta^-}\indup{tree}=
\bigeval{\exp\bigbrk{W_0(x,\check\Phi^+,\check\Phi^-)}\, \Op_\alpha^+\Op_\beta^- }_{\Phi=0},
\]
where $W_0$ is the tree-level Green function
\[\label{eq:W0}
W_0(x,\check\Phi^+,\check\Phi^-)=I_{0x}\Tr\check\Phi_m^+\check\Phi_m^-.
\]
The scalar propagator $I_{xy}$ is defined in \eqref{eq:prop}.
In order for the result to be non-vanishing 
all the scalar fields $\Phi^-$ in $\Op_\beta^-$
need to be contracted to fields $\Phi^+$ in $\Op_\alpha^+$ with 
the propagator $I_{0x}$.
In particular the number of fields of the two
operators must be equal.

\subsection{Renormalization}
\label{ssec:onelooprenorm}

For the one-loop correlator we insert the (connected) one-loop 
Green function $W_2$ into the tree-level correlator \eqref{eq:treelevelW}
\[\label{eq:oneloopW}
\bigvev{\Op_\alpha^+ \Op_\beta^-}\indup{one-loop}=
\bigeval{\exp\bigbrk{W_0(x,\check\Phi^+,\check\Phi^-)}
\bigbrk{1+\gnorm^2 W_2(x,\check\Phi^+,\check\Phi^-)}\Op_\alpha^+\Op_\beta^- }_{\Phi=0},
\]
where $\gnorm^2=\gym^2/16\pi^2$.
We now change the argument $\check \Phi^+$ of $W_2$ to $I^{-1}_{0x}\Phi^-$.
This can be done because the result vanishes unless
every $\Phi^-$ is contracted with some $\Phi^+$ 
before the fields $\Phi$ are set to zero.
Here, the only possibility is to contract with a term in
$W_0$ which effectively changes
the argument back to $\check \Phi^+$. In doing that we need to make sure
that no new contractions appear between the 
arguments $\Phi^-$ and $\check\Phi^-$ of $W_2$. 
Formally, this is achieved by `normal ordering' $\normord{\phantom{x}}$.
The correlator becomes
\[\label{eq:oneloopV}
\bigvev{\Op_\alpha^+ \Op_\beta^-}\indup{one-loop}=
\bigeval{\exp\bigbrk{W_0(x,\check\Phi^+,\check\Phi^-)}
\bigbrk{1+\gnorm^2 V_2^-(x)}\Op_\alpha^+ \Op_\beta^- }_{\Phi=0},
\]
with the one-loop effective vertex
\[\label{eq:V2}
V_2(x)=\normord{W_2(x,I_{0x}^{-1}\Phi,\check\Phi)}.\]
Instead of replacing $\check\Phi^+$ we could have replaced
$\check \Phi^-$. This shows that in 
\eqref{eq:oneloopV} $V_2^-$ is equivalent to $V_2^+$. 
In other words $V_2$ is self-adjoint 
with respect to the tree-level scalar product.

We renormalize the operators according to 
\[\label{eq:renorm2}
\tilde\Op=\bigbrk{1-\half \gnorm^2 V_2(x_0)}\Op,
\]
and find 
\[\label{eq:oneloopVrenorm}
\bigvev{\tilde\Op_\alpha^+ \tilde\Op_\beta^-}\indup{one-loop}=
\bigeval{\exp\bigbrk{W_0(x,\check\Phi^+,\check\Phi^-)}
\bigbrk{1+\gnorm^2 V_2^-(x)-
\gnorm^2 V_2^-(x_0)}\Op_\alpha^+ \Op_\beta^- }_{\Phi=0}.
\]
In the present renormalizable field theory in dimensional regularization
the dependence of $V_2$ on $x$ is fixed, we write
\[\label{eq:V2x}
V_2(x)=\frac{\Gamma(1-\epsilon)}{\bigabs{\half\mu^2 x^2}^{-\epsilon}} V_2.\]
We send the regulator to zero and find
\[\label{eq:oneloopD}
\lim_{\epsilon\to 0}\bigbrk{V_2(x)-V_2(x_0)}
=
\log \bigbrk{x_0^2/x^2}\,D_2,
\]
with 
\[\label{eq:D2limit}
D_2=-\lim_{\epsilon\to 0} \epsilon V_2.\]
The final answer for the renormalized correlator 
at $\epsilon=0$ is 
\[\label{eq:oneloopfinal}
\bigvev{\tilde\Op_\alpha^+ \tilde\Op_\beta^-}\indup{one-loop}=
\bigeval{\exp(W_0)
\exp\bigbrk{\log (x_0^2/x^2)\gnorm^2 D_2^-}\Op_\alpha^+ \Op_\beta^- }_{\Phi=0},
\]
in agreement with conformal field theory. $D_2$ is the one-loop correction
to the dilatation generator. Furthermore, the coefficient of the correlator
is given by its tree-level value as $D_2$ multiplies the logarithm only.
Although we are interested in correlators of renormalized operators
as on the left-hand side of \eqref{eq:oneloopfinal},
we can work with bare operators as on the right hand side 
of \eqref{eq:oneloopfinal}. We will not distinguish between the
two types of operators $\tilde\Op$ and $\Op$. It will be understood that an
operator in a correlator is renormalized.

\subsection{Evaluation of diagrams}
\label{ssec:oneloopdiag}

\begin{figure}\centering
\parbox{1.5cm}{\centering\includegraphics{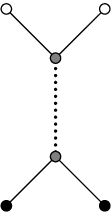}\par a}
\quad
\parbox{1.5cm}{\centering\includegraphics{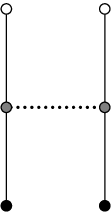}\par b}
\quad
\parbox{1.5cm}{\centering\includegraphics{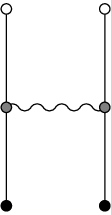}\par c}
\quad
\parbox{1.5cm}{\centering\includegraphics{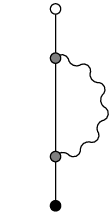}\par d1}
\quad
\parbox{1.5cm}{\centering\includegraphics{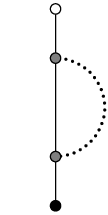}\par d2}
\quad
\parbox{1.5cm}{\centering\includegraphics{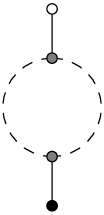}\par d3}
\quad
\parbox{1.5cm}{\centering\includegraphics{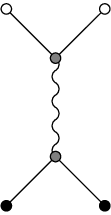}\par e}
\caption{One-loop Green functions.
The diagrams display the combinatorial structure with respect to the 
gauge group rather than their space-time configuration.
The solid, wiggly, dashed lines represent scalars, gluons, fermions, respectively.
The dotted lines correspond to a local interaction of four scalars/gluons.
}
\label{fig:oneloop}
\end{figure}%
The one-loop connected Green functions are 
depicted in \figref{fig:oneloop}.
To evaluate them we make use of 
the $\superN=4$ SYM action in \eqref{eq:SYMaction} with the 
coupling constant $\gnorm$ which is related to the usual coupling 
constant $\gym$ by 
$\gnorm^2=\gym^2/16\pi^2$.
The one-loop Green functions evaluate to 
\<\label{eq:W2abcd}
W_{2,a}\eq 
\quarter X_{00xx} 
\Tr\comm{\check\Phi^+_m}{\check\Phi^+_n}\comm{\check\Phi^-_m}{\check\Phi^-_n},
\nln
W_{2,b}\eq
\quarter X_{00xx} 
\bigbrk{\Tr\comm{\check\Phi^+_m}{\check\Phi^-_n}\comm{\check\Phi^+_m}{\check\Phi^-_n}
       +\Tr\comm{\check\Phi^+_m}{\check\Phi^-_n}\comm{\check\Phi^-_m}{\check\Phi^+_n}},
\nln
W_{2,c}\eq 
\bigbrk{-\half\tilde H_{0x,0x}-Y_{00x}I_{0x}+\quarter X_{00xx}}
\Tr\comm{\check\Phi^+_m}{\check\Phi^-_m}\comm{\check\Phi^+_n}{\check\Phi^-_n}
 ,
\nln
W_{2,d}\eq 
-Y_{00x}
\Tr\comm{\check\Phi^+_m}{T^a}\comm{T^a}{\check\Phi^-_m}.
\>
The functions $X,Y,\tilde H$ are defined in 
\eqref{eq:YXH}.
Diagram e vanishes by antisymmetry, its structure is
$\Tr\comm{\check\Phi^+_m}{\check\Phi^+_m}\comm{\check\Phi^-_n}{\check\Phi^-_n}=0$.
We use a Jacobi-Identity to transform the second structure
in $W_{2,b}$
\[\label{eq:Jacobi}
\Tr\comm{\check\Phi^+_m}{\check\Phi^-_n}\comm{\check\Phi^-_m}{\check\Phi^+_n}
=
\Tr\comm{\check\Phi^+_m}{\check\Phi^+_n}\comm{\check\Phi^-_m}{\check\Phi^-_n}
-\Tr\comm{\check\Phi^+_m}{\check\Phi^-_m}\comm{\check\Phi^+_n}{\check\Phi^-_n}.
\]
We shuffle some terms around and get 
\<\label{eq:W2ABC}
W_{2,A}\eq 
X_{00xx}
\bigbrk{\half\Tr\comm{\check\Phi^+_m}{\check\Phi^+_n}\comm{\check\Phi^-_m}{\check\Phi^-_n}
+\quarter\Tr\comm{\check\Phi^+_m}{\check\Phi^-_n}\comm{\check\Phi^+_m}{\check\Phi^-_n}},
\nln
W_{2,B}\eq 
-\half\tilde H_{0x,0x}
\Tr\comm{\check\Phi^+_m}{\check\Phi^-_m}\comm{\check\Phi^+_n}{\check\Phi^-_n}
 ,
\nln
W_{2,C}\eq 
-Y_{00x}
\bigbrk{I_{0x}\Tr\comm{\check\Phi^+_m}{\check\Phi^-_m}\comm{\check\Phi^+_n}{\check\Phi^-_n}
+\Tr\comm{\check\Phi^+_m}{T^a}\comm{T^a}{\check\Phi^-_m}}.
\>
According to \eqref{eq:V2} we write
\<\label{eq:V2ABC}
V_{2,A}\eq 
X_{00xx}I_{0x}^{-2}
\bigbrk{\half\normord{\Tr\comm{\Phi_m}{\Phi_n}\comm{\check\Phi_m}{\check\Phi_n}}
+\quarter\normord{\Tr\comm{\Phi_m}{\check\Phi_n}\comm{\Phi_m}{\check\Phi_n}}},
\nln
V_{2,B}\eq 
-\half\tilde H_{0x,0x}I_{0x}^{-2}
\normord{\Tr\comm{\Phi_m}{\check\Phi_m}\comm{\Phi_n}{\check\Phi_n}},
\nln
V_{2,C}\eq 
-Y_{00x}I_{0x}^{-1} \bigbrk{ 
\normord{\Tr\comm{\Phi_m}{\check\Phi_m}\comm{\Phi_n}{\check\Phi_n}}
+\normord{\Tr\comm{\Phi_m}{T^a}\comm{T^a}{\check\Phi_m}}}.
\>
The last term can now be written as
\<\label{eq:V2Cgauge}
V_{2,C}\eq
-Y_{00x}I_{0x}^{-1}\Tr\comm{\Phi_m}{\check\Phi_m}\comm{\Phi_n}{\check\Phi_n}
\nle
-Y_{00x}I_{0x}^{-1}\Tr\bigbrk{\comm{T^a}{\Phi_m}{\check\Phi_m}}
\Tr\bigbrk{\comm{T^a}{\Phi_n}{\check\Phi_n}}
\nle
-Y_{00x}I_{0x}^{-1} G^a G^a .
\>
The generator $G^a=\Tr\comm{T^a}{\Phi_m}{\check\Phi_m}$ 
is the generator of gauge transformations for scalar fields.
Therefore $V_{2,C}$ does not act on operators which are color 
singlets, it is irrelevant.
The remaining functions $X$ and $\tilde H$ have the following 
expansion in $\epsilon$,
see \eqref{eq:YXH2pt},
\<\label{eq:XHexpand}
\frac{X_{00xx}}{I_{0x}^2}\eq\lrbrk{\frac{2}{\epsilon}+2+\order{\epsilon^2}}
\frac{\Gamma(1-\epsilon)}{\bigabs{\half \mu^2x^2}^{-\epsilon}},\nln
\frac{\tilde H_{0x,0x}}{I_{0x}^2}\eq
\lrbrk{-48\zeta(3)\,\epsilon+\order{\epsilon^2}}
\frac{\Gamma(1-\epsilon)}{\bigabs{\half \mu^2x^2}^{-\epsilon}}.
\>
When inserted in \eqref{eq:D2limit}
this proves our initial statement about the one-loop
dilatation generator
\[\label{eq:D2A}
D_2=D_{2,A}=
-\normord{\Tr\comm{\Phi_m}{\Phi_n}\comm{\check\Phi_m}{\check\Phi_n}}
-\half\normord{\Tr\comm{\Phi_m}{\check\Phi_n}\comm{\Phi_m}{\check\Phi_n}}.
\]
%

\section{Two-loop}
\label{sec:twoloop}

In this chapter we will derive the central technical result of the
paper: We will find the two-loop correction to the dilatation
generator, as announced in eq.\eqref{introD4}. Written in 
$\grSO(6)$ notation, it is given by
\[\label{eq:D4}
D_4=-2 \normord{\Tr \Phi_m\comm{\Phi_n}{\comm{\check\Phi_m}{\comm{\Phi_p}{\comm{\check\Phi_n}{\check\Phi_p}}}}}
+\normord{\Tr \Phi_m\comm{\Phi_n}{\comm{T^a}{\comm{T^a}{\comm{\check\Phi_m}{\check\Phi_n}}}}}.
\]
This result is valid for \emph{pure} scalar operators.
Here pure means those composite operators, made from the six fundamental
scalar fields of the model, that do not mix with other, non-scalar
types of fields (or fields containing covariant derivatives). 
These pure scalars are given by the $\grSO(6)$ 
representations $[p,q,p]$ with $\Delta_0=2p+q$.
We start with a group-theoretical argument why the
dilatation generator closes on this class of operators.
We proceed by extending the renormalization scheme to two-loops.
Then we investigate the allowed structures in $D_4$ by considering
the relevant Feynman diagrams. Finally we fix the coefficients of 
the structures to obtain \eqref{eq:D4}.

\subsection{Group-theoretical constraints on mixing} 
\label{ssec:purescalars}

Restricting oneself to operators constructed from scalar fields
$\Phi_m$ only is 
in general not possible due to operator mixing:
Scalar operators can also contain scalar combinations of fermions $\Psi^A$, 
field strengths $F_{\mu\nu}$ and derivatives $\cder_\mu$.
All these operators are space-time singlets and 
they must be treated on equal footing. 
Nevertheless, one can isolate certain classes of operators 
by employing certain representations of the flavor symmetry
group $\grSO(6)\simeq\grSU(4)$. For example
the representations $(a,b,c)\simeq[b+c,a-b,b-c]$ (Young tableau/Dynkin labels) 
with classical scaling dimension $\Delta_0=a+b+|c|$ do not admit 
derivatives or field strengths. 
If furthermore $c=0$, fermions are excluded and the operator must consist 
of only scalar fields. We will therefore consider 
operators in the $\grSO(6)$ representation 
\[(a,b,0)\simeq[b,a-b,b]\qquad\mbox{with}\qquad \Delta_0=a+b.\]
These operators need not be superconformal primaries. 
If they are, they belong to the protected C series of
unitary irreducible representations (UIR) of $\grSU(2,2|4)$. 
In that case they are quarter BPS operators 
(for $b\neq 0$) or half BPS (for $b=0$).
Otherwise they belong the A series unprotected scalar UIR 
$(a-2,b-2,0)\simeq [b-2,a-b,b-2]$ with
$\Delta_0=a+b-2$. This UIR is at both unitarity bounds and splits
into four UIR's at $\gym=0$. The operators under consideration 
are primary operators of the highest-lying submultiplet. 

We will now prove the above two statements concerning absence of mixing.
The $\grSO(6)$ weights of the six scalars are $(\pm 1,0,0)$, $(0,\pm 1,0)$, 
$(0,0,\pm 1)$, the weights of the eight space-time pairs of fermions are 
$(\pm \half,\pm' \half,\pm'' \half)$ and derivatives as well as field strengths 
have trivial weight $(0,0,0)$. 
We see that the sum of labels $a+b+c$ for all of these is bounded from above 
by the corresponding scaling dimension of the field. 
For a composite operator the labels of the constituents simply are summed
and therefore $a+b+c$ is bounded from above by the scaling dimension $\Delta_0$. 
If this bound is to be achieved, all constituent fields must be on the bound
as well. This is true only for the three scalars with the weights
$(1,0,0)$, $(0,1,0)$, $(0,0,1)$ and a pair of fermions with 
weight $(\half,\half,\half)$. This proves the first statement. 
Furthermore if $c=0$ for the composite operator the third scalar and the fermions 
are excluded as constituents. This proves the second statement.

As advertised, an operator with $a+b=\Delta_0$, $c=0$ is constructed 
from two scalars only. These scalars are charged with respect to
different $\grU(1)$ subgroups of $\grSO(6)$. We choose
\[\label{eq:Zphi}
Z=\sfrac{1}{\sqrt{2}}\brk{\Phi_5+i\Phi_6},\quad
\phi=\sfrac{1}{\sqrt{2}}\brk{\Phi_1+i\Phi_2}.
\]
We identify $a,b$ with the number of $Z,\phi$ fields and
assume $a\geq b$. It may be shown that any combination of the fields
$Z,\phi$ is traceless with respect to the original $\grSO(6)$ symmetry. 
Written in terms of $Z,\phi$ the dilatation generator 
\eqref{eq:D0}, \eqref{eq:D2}, \eqref{eq:D4} is
\<\label{eq:D024Zphi}
D_0\eq
\Tr Z\check Z+\Tr \phi\check\phi,
\nln
D_2\eq
-2\,\normord{\Tr \comm{\phi}{Z}\comm{\check\phi}{\check Z}},
\nln
D_4\eq
-2\, \normord{\Tr \bigcomm{\comm{\phi}{Z}}{\check Z}\bigcomm{\comm{\check\phi}{\check Z}}{Z}}
\nl
-2\, \normord{\Tr \bigcomm{\comm{\phi}{Z}}{\check \phi}\bigcomm{\comm{\check\phi}{\check Z}}{\phi}}
\nl
-2\, \normord{\Tr \bigcomm{\comm{\phi}{Z}}{T^a}\bigcomm{\comm{\check\phi}{\check Z}}{T^a}}.
\>
For the derivation of the two-loop result, however, we
keep the $\grSO(6)$ symmetry manifest and consider the
correlator of two pure scalar $\grSO(6)$-traceless operators.

\subsection{Three pairs of legs}
\label{ssec:3legs}

\begin{figure}\centering
\parbox{2.0cm}{\centering\includegraphics{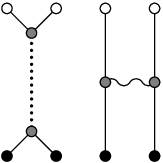}\par a}
\quad
\parbox{1.5cm}{\centering\includegraphics{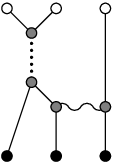}\par b}
\quad
\parbox{1.0cm}{\centering\includegraphics{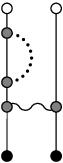}\par c}
\quad
\parbox{1.0cm}{\centering\includegraphics{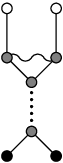}\par d}
\quad
\parbox{1.0cm}{\centering\includegraphics{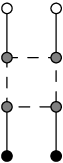}\par e}
\quad
\parbox{1.0cm}{\centering\includegraphics{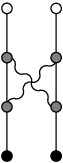}\par f}
\quad
\parbox{1.0cm}{\centering\includegraphics{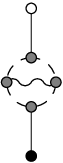}\par g}
\caption{Sample two-loop Green functions.}
\label{fig:twoloop}
\end{figure}%

At two loops the
interaction can be disconnected (\figref{fig:twoloop}a) or
connected. 
The disconnected diagrams can be shown to effectively yield 
$(\log)^2$ terms, necessary for the conformal structure
of the correlator, plus further connected diagrams, 
see \appref{sec:renorm}.
We classify the connected interactions by their number of legs. 
We can have three pairs of legs 
(\figref{fig:twoloop}b), two pairs of legs with one bulk loop 
(\figref{fig:twoloop}c,d,e,f)
or one pair of legs with two bulk loops (\figref{fig:twoloop}g). 
We start by analyzing the diagrams with three pairs of legs.

\begin{figure}\centering
\parbox{6cm}{\centering
\parbox{1.5cm}{\centering\includegraphics{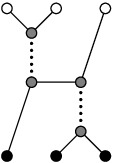}\par aa}
\parbox{1.5cm}{\centering\includegraphics{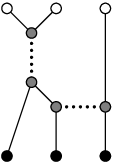}\par ab}
\parbox{1.5cm}{\centering\includegraphics{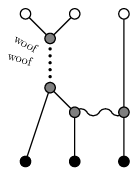}\par ac}\bigskip\par
\parbox{1.5cm}{\centering\includegraphics{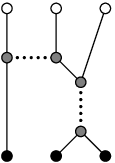}\par ba}
\parbox{1.5cm}{\centering\includegraphics{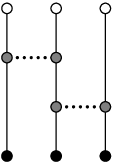}\par bb}
\parbox{1.5cm}{\centering\includegraphics{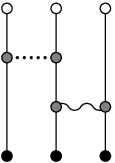}\par bc}\bigskip\par
\parbox{1.5cm}{\centering\includegraphics{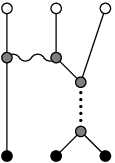}\par ca}
\parbox{1.5cm}{\centering\includegraphics{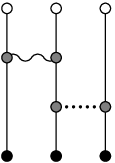}\par cb}
\parbox{1.5cm}{\centering\includegraphics{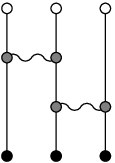}\par cc}}\qquad
\parbox{1.5cm}{\centering\includegraphics{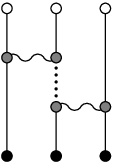}\par d1}
\parbox{1.5cm}{\centering\includegraphics{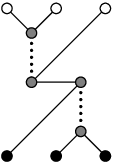}\par d2}
\parbox{1.5cm}{\centering\includegraphics{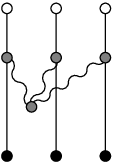}\par d3}
\caption{Two-loop Green functions with three pairs of legs.}
\label{fig:threelegs}
\end{figure}

The interactions with three pairs of legs are depicted in 
\figref{fig:threelegs}.
Most of the diagrams correspond to two one-loop interactions 
(see \figref{fig:oneloop})
connected at one leg. 
We start by subtracting the connected part of two 
one-loop vertices, see \eqref{eq:V4}.
This gives rise to the same structures but with different 
coefficient functions. 
As in the one-loop case the type b sub-diagrams
consist of two parts, see \eqref{eq:W2abcd}. The first part contains
local $\grSO(6)$ traces, $\check\Phi_m^+\check\Phi_m^+$. We have restricted 
ourselves to traceless operators and drop the term. The second
term can be transformed by a Jacobi identity to 
sub-diagrams of type a and c, see \eqref{eq:Jacobi}. This effectively
doubles the sub-diagrams of type a. 
Then we notice that
all type c sub-diagrams have a scalar line that is connected
to the rest of the diagram only by one gluon-line. 
By removing some normal ordering we can write those diagrams
as $G^a X^a$ plus diagrams with two pairs of legs
in analogy to \eqref{eq:V2Cgauge}. As $G^a$ is the generator of
gauge transformations and the operators are color singlets,
we can drop $G^a X^a$. The same holds for the d1 diagram.
The d2 diagram has local $\grSO(6)$ traces and will be dropped.
The d3 diagram (as well as a number of diagrams with
vertical gluon lines as in \figref{fig:oneloop}e, which have been
omitted in \figref{fig:threelegs}) 
vanishes by symmetry: 
The diagram is proportional to $\normord{G^a G^b G^c} \Tr \comm{T^a}{T^b}T^c$.

The only diagram with three pairs of legs that effectively
remains is \figref{fig:threelegs}aa. It is multiplied by four 
due to contributions from ab, ba and bb.
Its combinatorial structure is given by 
\[\label{eq:threelegs}
\normord{\Tr \Phi_m\comm{\Phi_n}{\comm{\check\Phi_m}{\comm{\Phi_p}{\comm{\check\Phi_n}{\check\Phi_p}}}}}.\]
%

\subsection{One and two legs}
\label{ssec:2legs}

There are many diagrams with one or two pairs of legs,
see \figref{fig:twoloop}c,d,e,f,g.
Additionally, the connected piece of two one-loop diagrams
and some reductions of diagrams with three pairs of legs 
have to be taken into account. Instead of considering
all these we investigate the possible algebraic structures. 
A generic diagram consists of one trace and four commutators. 
This can always be brought into the form
\[\label{eq:twoloopstructure}
\normord{\Tr \ldots\comm{\ldots}{\comm{\ldots}{\comm{\ldots}{\comm{\ldots}{\ldots}}}}}.\]
For diagrams with two pairs of legs we need to 
distribute $\Phi_m$, $\Phi_n$, $\check \Phi_m$, $\check\Phi_n$ and two $T^a$
on the six slots. We start by distributing the $T^a$.
If both $T^a$ are in the first two
or the last two slots, the structure vanishes by antisymmetry.
If we place a $T^a$ in the first or the last slot, we can interchange it 
with the second or second-but-last. 
There are identities when the two $T^a$ are next to each other
(corresponding to \figref{fig:twoloop}c)
\[\label{eq:Tnextto}
\comm{T^a}{\comm{T^a}{\comm{X}{Y}}}=
\comm{X}{\comm{T^a}{\comm{T^a}{Y}}}=2\lrbrk{\Tr T^aT^a/\Tr 1}\comm{X}{Y},
\]
or one step apart
(corresponding to \figref{fig:twoloop}d)
\[\label{eq:Tclose}
\comm{T^a}{\comm{X}{\comm{T^a}{Y}}}
=
\lrbrk{\Tr T^aT^a/\Tr 1}\comm{X}{Y}.
\]
The combination $\Tr T^aT^a/\Tr 1$ 
depends only on the gauge group, for $\grU(N)$ it equals $N$.
With these we can remove two commutators along with the $T^a$ and get two
independent structures
\[\label{eq:twolegsa}
\Tr \Phi_m\comm{\Phi_n}{\comm{T^a}{\comm{T^a}{\comm{\check\Phi_m}{\check\Phi_n}}}}
\sim
N\Tr\comm{\Phi_m}{\Phi_n}\comm{\check\Phi_m}{\check\Phi_n},
\]
and
\[\label{eq:twolegsb}
\normord{\Tr \Phi_m\comm{\check\Phi_m}{\comm{T^a}{\comm{T^a}{\comm{\Phi_n}{\check\Phi_n}}}}}
\sim
N\normord{\Tr\comm{\Phi_m}{\check\Phi_m}\comm{\Phi_n}{\check\Phi_n}}.
\]
By use of gauge invariance,
see \eqref{eq:V2Cgauge}, we can write \eqref{eq:twolegsb}
as a one pair of legs diagram.
The remaining structures have the two $T^a$ two steps apart
(corresponding to \figref{fig:twoloop}e,f).
We write them as
\[\label{eq:twolegsloop}
\normord{\Tr T^a\comm{X_1}{\comm{X_2}{\comm{X_3}{\comm{X_4}{T^a}}}}}.\]
Using Jacobi identities on the nested commutators it is easy to see
that we can interchange any of the $X_i$ at the cost of 
terms like \eqref{eq:Tclose}. 
The only remaining independent structure can thus be chosen as 
\[\label{eq:twolegsc}
\normord{\Tr T^a\comm{\Phi_m}{\comm{\check \Phi_m}{\comm{\Phi_n}{\comm{\check\Phi_n}{T^a}}}}}.\]

By use of the identities \eqref{eq:Tnextto} and \eqref{eq:Tclose} 
a diagram with one pair of legs (\figref{fig:twoloop}g) can always be written as
\[\label{eq:oneleg}
\Tr \Phi_m\comm{T^a}{\comm{T^a}{\comm{T^b}{\comm{T^b}{\check\Phi_m}}}}\sim
N^2\Tr \Phi_m\check\Phi_m-N\Tr \Phi_m\Tr\check\Phi_m.
\]
%

\subsection{Determining the coefficients}
\label{ssec:twoloopcoeff}

All in all there are four independent structures
\eqref{eq:threelegs}, \eqref{eq:twolegsa},
\eqref{eq:twolegsc} and \eqref{eq:oneleg}.
The dilatation generator $D_4$ at two-loops must be a linear combination of these.
By using known results we determine this linear combination.

We start by applying the structures to the half BPS operator
$\Tr Z^J$. This operator is protected, it does not receive
corrections to its scaling dimension. Consequently $D_4$
must annihilate it. Indeed the structures \eqref{eq:threelegs} and
\eqref{eq:twolegsa} do so. The structure \eqref{eq:oneleg}
gives $JN^2 \Tr Z^J$ while \eqref{eq:twolegsc} gives
\[\label{eq:invalidoperation}
2J\sum_{p=1}^{J-1}\bigbrk{N\Tr Z^p \Tr Z^{J-p}+\Tr Z^2\Tr Z^{p-1}
\Tr Z^{J-1-p}},
\]
(up to terms involving $\Tr Z$). Thus
in order for $D_4$ to annihilate $\Tr Z^J$ 
the coefficients of \eqref{eq:oneleg} and \eqref{eq:twolegsc}
must vanish. 
This vanishing was explicitly confirmed
in references~\cite{Penati:1999ba,Penati:2000zv}. 
Our two independent structures correspond directly to those appearing there.

We are now left with
\[\label{eq:twoloopstruct}
D_4=\alpha \normord{\Tr \Phi_m\comm{\Phi_n}{\comm{\check\Phi_m}{\comm{\Phi_p}{\comm{\check\Phi_n}{\check\Phi_p}}}}}
+\beta \Tr \Phi_m\comm{\Phi_n}{\comm{T^a}{\comm{T^a}{\comm{\check\Phi_m}{\check\Phi_n}}}}.
\]
For the BMN-operators, see \secref{sec:BMNlimit}, 
\eqref{eq:twoloopstruct} produces the two-loop planar anomalous dimension
\[\label{eq:BMNdimpar}
-\frac{(2\beta+\alpha)\gym^4N^2n^2}{16\pi^2J^2}
+\frac{\alpha\gym^4N^2n^4}{8J^4}.
\]
By the pp-wave/BMN correspondence \cite{Berenstein:2002jq}
we would expect the first term to vanish 
in order for the BMN limit
\eqref{BMNlim} to be well-defined. The second term should equal
\[\label{eq:BMNdimres}
-\quarter\lambda^{\prime\,2}n^4=
-\frac{\gym^4N^2n^4}{4J^4}.
\]
This is also the result of an explicit calculation 
\cite{Gross:2002su} and an investigation 
using superconformal symmetry \cite{Santambrogio:2002sb}.
We find 
\[\label{eq:BMNdimfit}
\alpha=-2,\quad \beta=1.\]
This concludes our derivation of the two-loop dilatation generator
\eqref{eq:D4}.

\section{Applications to lower dimensional operators}
\label{sec:lowdim}

In this section we will apply the two-loop dilatation generator
to a set of lower dimensional operators.
We consider only those representations where mixing 
with fermionic and derivative insertions is 
prohibited, see \secref{ssec:purescalars}.
More explicitly, these are space-time scalars
in the $\grSO(6)$ representations $[p,\Delta_0-2p,p]$.
Furthermore, we will consider $\grSU(N)$ as the gauge
group. This is equivalent to dropping
all traces of single fields. For low dimensions the calculations
are easily performed by hand; as the dimension increases it
is useful to use an appropriate code in order 
to quickly and reliably work out the
action of the dilatation operator%
\footnote{Many results of this section and 
\secref{sec:highergrounds} have been obtained 
using \texttt{Mathematica}. 
Computing e.g. the exact dilatation matrix at two-loops 
for the 48 operators of dimension 8 took 90 sec.\ 
on a 366 MHz PC.
Obviously it is also useful and straightforward
to use \texttt{Mathematica} or similar
to diagonalize the dilatation matrices and perform the $\frac{1}{N}$
expansion of the eigensystem.}.

\subsection{Quarter BPS operators}
\label{ssec:quarterBPS}

Quarter BPS operators belong to the representations $[p,q,p]$ with
$p\geq1 $ and have classical scaling dimension $\Delta_0=2p+q$. 
(For $p=0$ the operators are half BPS.) 
In reference~\cite{Ryzhov:2001bp} the explicit construction of lower
dimensional $1/4$ BPS operators was initiated. The strategy consisted in
first writing down all linearly independent local polynomial scalar
composite operators in a given representation and next diagonalizing the
corresponding matrix of one-loop two-point functions in the inner product
given by the matrix of tree-level two-point functions. The former part of
the strategy was very recently
 systematized and completed in~\cite{D'Hoker:2003vf}.
Our idea of focusing on the dilatation operator of the theory allows us to
significantly simplify and extend the analysis
of~\cite{Ryzhov:2001bp,D'Hoker:2003vf}. As explained in
\secref{sec:intro} one does not need to work with the tree-level
inner product but can by use of the effective vertices~\eqref{eq:D2A}
and~\eqref{eq:D4} directly write down the action of the dilatation
generator in any convenient basis of operators. The (possible) 1/4 BPS
operators are then simply determined as the kernel of the dilatation
operator. 
\begin{table}\centering
\begin{tabular}[t]{|c|c||ccc|}\hline
dim & $\grSO(6)$  & BPS & $+$ & $-$
\\\hline\hline
2&[0,2,0]&1&-&-
\\\hline
3&[0,3,0]&1&-&-
\\\hline
4&[0,4,0]&2&-&-\\
 &[2,0,2]&1&1&-
\\\hline
5&[0,5,0]&2&-&-\\
 &[1,3,1]&1&-&-\\
 &[2,1,2]&1&1&-
\\\hline
6&[0,6,0]&4&-&-\\
 &[1,4,1]&1&-&-\\
 &[2,2,2]&3&3&-\\
 &[3,0,3]&-&-&1
\\\hline
7&[0,7,0]&4&-&-\\
 &[1,5,1]&3&-&-\\
 &[2,3,2]&4&4&-\\
 &[3,1,3]&2&2&1
\\\hline
\end{tabular}\qquad
\begin{tabular}[t]{|c|c||ccc|}\hline
dim & $\grSO(6)$ & BPS & $+$ & $-$
\\\hline\hline
8&[0,8,0]&7&-&-\\
 &[1,6,1]&4&-&-\\
 &[2,4,2]&8&8&-\\
 &[3,2,3]&3&4&3\\
 &[4,0,4]&4&7&-
\\\hline
9&[0,9,0]&8&-&-\\
 &[1,7,1]&7&-&-\\
 &[2,5,2]&11&11&-\\
 &[3,3,3]&9&11&5\\
 &[4,1,4]&5&11&2
\\\hline
10&[0,10,0]&12&-&-\\
 &[1,8,1]&10&-&-\\
 &[2,6,2]&19&19&-\\
 &[3,4,3]&13&18&9\\
 &[4,2,4]&15&33&5\\
 &[5,0,5]&1&4&8
\\\hline
\end{tabular}
\caption{BPS and unprotected operators up to dimension $10$.
The unprotected operators of representation $\Delta_0$, $[p,q,p]$
belong to the $\grSU(2,2|4)$ UIR $[p-2,q,p-2]$ of bare dimension
$\Delta_0-2$.
They are separated into two groups, $+$ and $-$
according to their parity, \emph{cf}.\ \secref{sec:NoMystery}.}
\label{tab:quarterBPS}
\end{table}
By our method we have confirmed the BPS nature of the operators
found in~\cite{Ryzhov:2001bp,D'Hoker:2003vf} up to two-loop order
(without any  modification of mixing coefficients 
at $\order{\gym^2}$).
Furthermore, we have determined all 1/4 BPS operators of dimension 8 (to
two-loop order) and of dimension 9 and 10 (one-loop order). Needless to
say that our method also gives access to a host of previously unknown
two-loop anomalous dimensions of unprotected operators. Our results on BPS
operators are summarized in~\tabref{tab:quarterBPS} and a collection of
anomalous dimensions can be found in the following sections 
and in \appref{sec:anodim}. 
It is known that operators in the representations $[1,p,1]$ are always 1/4 BPS.
For the operators considered in \tabref{tab:quarterBPS} it appears that
exactly half of those belonging to representations of the type $[2,p,2]$
are 1/4 BPS, see also \secref{sec:twoimp}. 

In the following two sub-sections we shall discuss some further
insights gained from our analysis.

\subsection{Exact results for lower dimensional operators}
\label{ssec:Dim234}

{}From \tabref{tab:quarterBPS} we  see that there
is exactly one unprotected operator in 
each of the representations 
($\Delta_0=4$, $[2,0,2]$), 
($\Delta_0=5$, $[2,1,2]$) and
($\Delta_0=6$, $[3,0,3]$). 
These operators can be written as
\<\label{eq:singleops}
\mathcal{K}'=\Op\indup{a}\eq\Tr \comm{\phi}{Z}\comm{\phi}{Z},
\nln
\Op\indup{b}\eq\Tr \comm{\phi}{Z} \comm{\phi}{Z}Z,
\nln
\Op\indup{c}\eq\Tr \comm{\phi}{Z} \comm{\phi}{Z}\comm{\phi}{Z},
\>
and descend from the operators
\<\label{eq:parents}
\mathcal{K}=\Op\indup{a,0}\eq\Tr \Phi_m\Phi_m,
\nln
\Op\indup{b,0}\eq\Tr \Phi_m\Phi_m \Phi_n,
\nln
\Op\indup{c,0}\eq\Tr \Phi_m\Phi_m \comm{\Phi_n}{\Phi_p},
\>
with
($\Delta_0=2$, $[0,0,0]$), 
($\Delta_0=3$, $[0,1,0]$) and
($\Delta_0=4$, $[1,0,1]$). 
For the operators~\eqref{eq:singleops}
the dilatation generator 
yields
\<\label{eq:singledim}
\Delta_{\mathcal{K}'}=\Delta\indup{a}\eq 4+\frac{3 \gym^2 N}{4\pi^2}
-\frac{3 \gym^4 N^2}{16\pi^4}, 
\nln
\Delta\indup{b}\eq 5+\frac{\gym^2N}{2\pi^2}
-\frac{3 \gym^4 N^2}{32 \pi^4}, 
\nln
\Delta\indup{c}\eq 6+\frac{3\gym^2N}{4\pi^2}
-\frac{9 \gym^4 N^2}{64 \pi^4},
\>
and these values do not receive any non-planar corrections at the 
present
loop order. 
Note that the anomalous dimension of the Konishi descendant $\mathcal{K}'$
agrees with previous results~\cite{Bianchi:1999ge,Bianchi:2000hn,Bianchi:2001cm}
and that $\Delta_{\mathcal{K}'}$ and $\Delta\indup{b}$ are in agreement
with the general formula~\eqref{andim}, corresponding respectively
to the cases $(J,n)=(2,1)$ and $(J,n)=(3,1)$.
The operator $\Op\indup{c}$ is the first in an infinite
sequence of operators with identical quantum anomalous dimensions, to be
discussed in \secref{ssec:3tower}.

\subsection{Peculiarities of the large $N$ expansion}
\label{ssec:degen}

Most of the other single-trace
operators in \tabref{tab:quarterBPS}
cease to be 
eigenvectors of the one-loop dilatation generator when we include non-planar
corrections, due to mixing with multiple trace operators. This mixing
has important consequences for the very nature of the anomalous dimension
of these operators as we shall illustrate by the examples
of ($\Delta_0=6$, $[2,2,2]$) and ($\Delta_0=7$, $[2,3,2]$) from 
\tabref{tab:quarterBPS}.

The operators in the first example
are descendants of the operators 
with ($\Delta_0=4$, $[0,2,0]$) studied in \cite{Bianchi:2002rw}.
The three unprotected operators can be chosen as
\[\label{J4ops}
\Op=\left(\begin{array}{c}
\Tr \comm{\phi}{Z}\comm{\phi}{Z}Z^2\\
\Tr \comm{\phi}{Z}Z\comm{\phi}{Z}Z\\\hline
\Tr Z^2\Tr \comm{\phi}{Z}\comm{\phi}{Z}
\end{array}\right),
\]
the commutators conveniently projecting out the three protected ones. 
In the basis given by~\eqref{J4ops} the one- and two-loop dilatation 
generator take the form
\[\label{D2J4}
D_2=
N\left(\begin{array}{cc|c}
+12&-2&+\sfrac{6}{N}\\
-8&+8&-\sfrac{4}{N}\\\hline
+\sfrac{16}{N}&-\sfrac{16}{N}&+12
\end{array}\right),
\]
and
\[\label{D4J4}
D_4=
N^2
\left(\begin{array}{cc|c}
-52-\sfrac{24}{N^2}&+4+\sfrac{24}{N^2}&-\sfrac{52}{N}\\
+48+\sfrac{16}{N^2}&-16-\sfrac{16}{N^2}&+\sfrac{48}{N}\\\hline
-\sfrac{144}{N}&+\sfrac{80}{N}&-48-\sfrac{64}{N^2}
\end{array}\right).
\]

Similarly, for $\Delta_0=7$ a convenient basis of unprotected operators
consists of the set of operators
given by
\[\label{J5ops}
\Op=\left(\begin{array}{c}
\Tr \comm{\phi}{Z}\comm{\phi}{Z}Z^3\\
\Tr \comm{\phi}{Z}Z\comm{\phi}{Z}Z^2\\\hline
\Tr Z^2\Tr \comm{\phi}{Z}\comm{\phi}{Z}Z\\
\Tr Z^3\Tr \comm{\phi}{Z}\comm{\phi}{Z}
\end{array}\right),
\]
and in this basis the one- and two-loop dilatation matrices read
\[\label{D2J5}
D_2=N
\left(\begin{array}{cc|cc}
+12&0&0&+\sfrac{6}{N}\\
-4&+4&+\sfrac{4}{N}&-\sfrac{6}{N}\\\hline
+\sfrac{8}{N}&-\sfrac{8}{N}&+8&0\\
+\sfrac{24}{N}&-\sfrac{24}{N}&0&+12
\end{array}\right),
\]
and
\[\label{D4J5}
D_4=N^2
\left(\begin{array}{cc|cc}
-48-\sfrac{36}{N^2}&-12+\sfrac{36}{N^2}&0&-\sfrac{48}{N}\\
+20+\sfrac{4}{N^2}&-\sfrac{4}{N^2}&-\sfrac{16}{N}&+\sfrac{20}{N}\\\hline
-\sfrac{54}{N}&+\sfrac{16}{N}&-24+\sfrac{8}{N^2}&-\sfrac{24}{N^2}\\
-\sfrac{192}{N}&+\sfrac{72}{N}&+\sfrac{24}{N^2}&-48-\sfrac{72}{N^2}
\end{array}\right).
\]
Notice that the dilatation matrices~\eqref{D2J4}, \eqref{D4J4}, \eqref{D2J5}
and~\eqref{D4J5} are exact to the given loop
order. Not surprisingly, these matrices have a form compatible with
the existence of a 't Hooft expansion, trace splitting and trace
joining processes being suppressed by a power of $\frac{1}{N}$ compared to 
trace conserving ones. 

At the classical level all the operators in respectively \eqref{J4ops} and 
\eqref{J5ops} have conformal dimension respectively 6 and 7. 
Including quantum effects at one-loop
and two-loop order the proper scaling operators consist
of some operators which are mainly single trace and some which are 
mainly multiple trace. The corresponding quantum corrections to the
conformal dimensions are the 
roots of the characteristic polynomial
of the matrix $\gnorm^2D_2+\gnorm^4D_4$. It is easy to show that this
characteristic polynomial has a well-defined 't Hooft expansion with
the dependence of $\gym^2$ and $N$ organizing into the 't Hooft coupling
$\lambda=\gym^2N$ and the genus counting parameter $\frac{1}{N^2}$.
Furthermore, its topological expansion terminates at a finite genus.
However, the \emph{roots} of
the characteristic polynomial need not have a terminating genus
expansion%
\footnote{This was also 
noticed in~\cite{Ryzhov:2001bp,Bianchi:2002rw,Arutyunov:2002rs}.}
or even a well defined 't Hooft expansion. 
Let us illustrate
this by considering in more detail the operators
\eqref{J4ops}, \eqref{J5ops}.

For $\Delta_0=6$ the relevant characteristic polynomial is of cubic order
\<
&&\omega^3
- 8\omega^2 
+ \lrbrk{20-\frac{10}{N^2}}\omega 
- \lrbrk{15-\frac{10}{N^2}}
\\\nn&&\phantom{\omega^3}
+\frac{\gym^2N}{16\pi^2}
\lrbrk{
\lrbrk{29+\frac{26}{N^2}}\omega^2 
- \lrbrk{141-\frac{54}{N^2}}\omega
+ \lrbrk{150-\frac{100}{N^2}}
},
\>
and the scaling dimensions up to two-loops are exactly 
\[
\Delta=6+\frac{\gym^2N}{4\pi^2}\,\omega,
\]
with $\omega$ a root of the corresponding cubic equation.
The discriminant of the cubic equation does not take
the form of a perfect square but is easily seen to be expressible as
\[
{\cal D}=\sum_{i,j} c_{i,j}\, \frac{(g^2 N)^i}{N^{2j}}, 
\]
with the $c_{i,j}$ some constants and $c_{0,0}\neq 0$, implying that
the roots will have a well-defined 't Hooft expansion which however involves
a non-terminating sum over genera. To three leading orders in the genus
expansion the conformal dimension of the two mainly single trace operators
read
\<\label{eq:BianchiSingle}
\Delta\indup{single}\eq6+
\frac{\gym^2N}{\pi^2}\lrbrk{
\frac{5\pm\sqrt{5}}{8}
+\frac{5\pm2\sqrt{5}}{2N^2}
-\frac{75\pm34\sqrt{5}}{N^4}} 
\\\nn
&&\mathrel{}\mathord{\phantom{6}}+\frac{\gym^4N^2}{\pi^4}\lrbrk{
-\frac{17\pm5\sqrt{5}}{128}
-\frac{131\pm57\sqrt{5}}{64N^2}
+\frac{1675\pm751\sqrt{5}}{16N^4}},
\>
whereas that of the mainly double trace one equals
\[\label{eq:BianchiDouble}
\Delta\indup{double}=6+\frac{\gym^2 N}{\pi^2}\lrbrk{\frac{3}{4}-\frac{5}{N^2}+\frac{150}{N^4}}
+\frac{\gym^4 N^2}{\pi^4}\lrbrk{-\frac{3}{16}
+\frac{59}{16N^2}-\frac{1675}{8N^4}}.
\nn
\]
The one-loop contributions to these conformal dimensions agree with
those obtained in reference~\cite{Bianchi:2002rw} and the planar
versions of $\Delta\indup{single}$ are in accordance with the general 
formula~\eqref{andim} for $J=4$ and $n=1$ and $n=2$ respectively.
Furthermore the planar part of the quantum correction to $\Delta\indup{double}$
is identical to that of the operator $\mathcal{K}'$ in~\eqref{eq:singleops}
as it should be.

In the case $\Delta_0=7$ the situation is more involved. 
Here the relevant
characteristic polynomial is of degree four. The
discriminant of the associated cubic polynomial does also not here
take the form of a perfect square but can be expressed as
\[\label{disexp}
{\cal D}=\sum_{i,j} c_{i,j}\, \frac{(g^2 N)^i}{N^{2j}},
\]
where the $c_{i,j}$ are again some constants. However, this time
{$c_{0,0}=c_{1,0}=0$}. 
This implies that not all 
the wished for conformal
dimensions have well defined 't Hooft expansions. 
(Of course they still exist as the exact roots of the characteristic
polynomial.)
Two of the conformal dimensions, associated with respectively a mostly
single trace and a mostly double trace operator, do have well defined 
't Hooft expansions and for these we find for the first few orders
\<
\Delta\indup{single}\eq
7+\frac{\gym^2N}{\pi^2}
\lrbrk{\frac{1}{4}+\frac{11}{16N^2}+\frac{253}{64N^4}}
+\frac{\gym^4N^2}{\pi^4}
\lrbrk{-\frac{3}{128}-\frac{103}{1024N^2}-\frac{5541}{8192N^4}},
\nln
\Delta\indup{double}\eq
7+\frac{\gym^2N}{\pi^2}
\lrbrk{\frac{1}{2}-\frac{1}{2N^2}-\frac{17}{2N^4}}
+\frac{\gym^4N^2}{\pi^4}
\lrbrk{-\frac{3}{32}+\frac{7}{32N^2}+\frac{129}{32N^4}}.
\>
Here we notice that the planar contribution to $\Delta\indup{single}$ is in
accordance with the general formula~\eqref{andim} for $(J,n)=(5,1)$
and that the planar quantum correction to $\Delta\indup{double}$ 
as expected is the same as that of the operator $\mathcal{O}_b$ in 
equation~\eqref{eq:singleops}.
 We then remove the roots 
associated to these scaling dimensions and obtain from the quartic polynomial
a quadratic one. This quadratic polynomial has the roots 
\<\label{J5weird}
\Delta\indup{remaining}\eq7+\frac{\gym^2 N}{\pi^2}\bigbrk{\mathcal{M}\pm \sqrt{\mathcal{D}}},
\\
\mathcal{M}\eq
\lrbrk{\frac{3}{4}-\frac{3}{32N^2}+\frac{291}{128N^4}}
+\frac{\gym^2 N}{\pi^2}\lrbrk{-\frac{45}{256}-\frac{537}{2048N^2}-\frac{27483}{16384N^4}},
\nln\nn
\mathcal{D}\eq
\lrbrk{\frac{27}{32N^2}-\frac{531}{1024N^4}}
+\frac{\gym^2 N}{\pi^2}\lrbrk{-\frac{3123}{4096N^2}+\frac{29295}{32768N^4}}
+\frac{9\gym^4 N^2}{65536\pi^4}.
\>
As we have determined the two well-defined scaling dimensions in 
a perturbative fashion, the coefficients $\mathcal{M},\mathcal{D}$ 
of the new polynomial are series in $\frac{1}{N}$ and $\gym^2N$.
(Here we notice that the last coefficient in $\mathcal{D}$ is indeed determined
by $D_2$ and $D_4$ alone and as we shall see it plays a crucial role for the
analysis.) {}From the form of $\mathcal{D}$ it follows that we have a 
square
root singularity in the $(\lambda,\frac{1}{N})$ plane with a branch point
at the origin. Obviously, we can not expand $\sqrt{\mathcal{D}}$ in
powers of $\lambda$ and $\frac{1}{N}$ at the same time. Thus
for the remaining two operators it only makes sense to speak
of either a singular anomalous dimension \eqref{J5weird},
a planar anomalous dimension ($N\rightarrow \infty$) 
or a one-loop anomalous dimension ($\gnorm\rightarrow 0$). For the
planar anomalous dimension we find 
\<\label{planarJ5}
\Delta\indup{single,planar}\eq
7+\frac{3\gym^2N}{4\pi^2}-\frac{21\gym^4N^2}{128\pi^4}, 
\nln
\Delta\indup{double,planar}\eq
7+\frac{3\gym^2N}{4\pi^2}-\frac{3\gym^4N^2}{16\pi^4}.
\>
The first of these anomalous dimensions correctly reproduces the result
encoded in the general formula~\eqref{andim} and the quantum correction
of the second agrees with that of the Konishi operator $\mathcal{K}$.
For the one-loop anomalous dimension one gets to four leading orders
in the large-$N$ expansion
\[\label{oneloopJ5}
\Delta\indup{one-loop}=7+
\frac{\gym^2N}{\pi^2}\lrbrk{\frac{3}{4}\pm\frac{3\sqrt{6}}{8N}-\frac{3}{32N^2}
\mp\frac{59\sqrt{6}}{512N^3}}.
\] 
The absence of a well-defined 't Hooft expansion for the conformal dimension
of certain operators is a degeneracy effect and the appearance of odd powers
of $\frac{1}{N}$ is a particular manifestation of this. 
{}From the form of the one-loop dilatation matrix~\eqref{D2J5}
it follows that at the
planar level a single trace state and a double trace state
are degenerate. Treating the $\frac{1}{N}$ effects as a perturbation
thus calls for the use of \emph{degenerate} perturbation theory which
exactly gives rise to the $\frac{1}{N}$ corrections 
in equation~\eqref{oneloopJ5}, lifting the degeneracy present at the one-loop
level. The degeneracy is also lifted if instead the two-loop effects are
treated perturbatively and it is this simultaneous lift of degeneracy by
two non-commuting perturbations which is the origin of the problems with
the 't Hooft expansion. Note that these problems also renders problematic
the perturbative evaluation of higher point functions in conjunction
with the $\frac{1}{N}$ expansion. This question is however beyond the 
scope of this paper.

\section{Two impurity operators}
\label{sec:twoimp}

Berenstein, Maldacena and Nastase suggested to 
investigate operators of a large dimension $\Delta_0$
and a nearly equally large charge $J$ under a subgroup 
$\grSO(2)$ of $\grSO(6)$. 
We identify the $\grSO(2)$ subgroup with the group $\grU(1)$ 
corresponding to the phase of the field $Z$ defined in \eqref{eq:Zphi}.
Then the relevant operators constitute long strings of 
$Z$-fields with $\Delta_0-J$ impurities scattered in. 
As particular examples BMN investigated operators 
with zero, one and two impurities of type $\phi$. 
As shown in \cite{Beisert:2002tn} it makes perfect sense
to consider these operators also for arbitrary finite values of $J$. 
The primary single-trace operators with up to two impurities are 
in one-to-one correspondence
with the $\grSO(6)$ representations 
$[0,\Delta_0,0]$ and $[0,\Delta_0-2,0]$.
These primary operators have supersymmetry 
descendants that can be written in terms 
of $J$ fields $Z$ and two impurities of type $\phi$.
They belong to the representation $[2,\Delta_0-4,2]$ 
for which we may apply the dilatation generator up to two-loops
as given by~\eqref{eq:D024Zphi}.
A generic multi-trace operator with two impurities is 
written as%
\footnote{We observe that there seem to be as many unprotected 
operators of the type \eqref{symop} as there are BPS operators of the type \eqref{bpsop} in 
the $\grSO(6)$ representation $[2,\Delta_0-4,2]$,
see also \tabref{tab:quarterBPS}.}
\<\label{symop}
\Op_p^{J_0;J_1,\ldots,J_k}\eq \Tr(\phi Z^p \phi Z^{J_0-p})
\prod_{i=1}^k \Tr Z^{J_i},
\\
\label{bpsop}
\OpV^{J_0,J_1;J_2,\ldots,J_k}\eq
\Tr(\phi Z^{J_0})\Tr(\phi Z^{J_1})
\prod_{i=2}^k \Tr Z^{J_i},
\>
with $\sum_{i=0}^k J_i=J$.
Both series of operators are symmetric under the interchange of 
sizes $J_k$ of traces $\Tr Z^{J_k}$, $\Op$ is symmetric under
$p\to J_0-p$ and $\OpV$ is symmetric under $J_0\leftrightarrow J_1$.

The quantum dilatation generator~\eqref{eq:D024Zphi} 
can be seen to act as
\[
(g^2D_2+g^4D_4)\matr{c}{\Op\\\OpV}=
\matr{cc}{\ast&0\\\ast&0}\matr{c}{\Op\\\OpV},
\]
i.e. operators of type $\OpV$ are never produced. 
This follows from the fact that all produced 
objects will contain a commutator $\comm{\phi}{Z}$ 
in some trace and this trace
will vanish unless it contains another $\phi$.
It immediately shows that for every $\OpV$ there is 
one protected (quarter BPS) operator.
Its leading part is given by $\OpV$ itself, plus a $\frac{1}{N}$ 
correction from the operators $\Op$ 
\cite{Ryzhov:2001bp,D'Hoker:2003vf,Beisert:2002bb,Constable:2002vq}. 
On the other hand the operators $\Op$ are in general not protected
and we will investigate their spectrum of anomalous dimensions in 
the following. {}From the form of the dilatation matrix we 
infer that operators of type $\Op$ 
do not receive corrections
from operators of type $\OpV$; the latter therefore completely decouple
as far as the consideration of the $\Op$'s is concerned.

\subsection{The action of the dilatation generator}
\label{ssec:bmn}

When acting on states of the type given in~\eqref{symop}
the one- and two-loop dilatation operator
take the form given in equation~\eqref{eq:D024Zphi}, i.e.
\<
D_2\eq-2\,\normord{\Tr [Z,\phi][\check{Z},\check{\phi}]},
\nln\label{Z2}
D_4\eq 2\,\normord{\Tr[Z,\phi][\check{Z},[Z,[\check{Z},\check{\phi}]]]}
\nlnum\label{Noneloop}
+4N\,\normord{\Tr [Z,\phi][\check{Z},\check{\phi}]} 
\nlnum
+2\,\normord{\Tr[Z,\phi][\check{\phi},[\phi,[\check{Z},\check{\phi}]]]}. 
\nn 
\>
Here we shall prove a further simplification occurring when $D_4$ acts on
\eqref{symop}, namely
\<
D_4\eq
-\quarter \left(D_2\right)^2 
+2\,\normord{\Tr[Z,\phi][\check{\phi},[\phi,[\check{Z},\check{\phi}]]]}
\label{square} 
\\&\equiv&\mathrel{}
-\quarter\left(D_2\right)^2 +\delta D_4.
\label{deltaD}
\>
This relation explicitly shows that $D_2$ and $D_4$ do not commute
due to the extra piece $\delta D_4$, involving an interaction of
both impurities that was not present at the one-loop level.
Eq. \eqref{square} is shown as follows: When applying $D_2$ to an 
operator $\Op_p$ of the type~\eqref{symop}
there are two possible ways of contracting
the $\check{\phi}$ in $D_2$ with
 a $\phi$ in $\Op_p$. Each of these possibilities
gives the same contribution to $D_2\Op_p$
\footnote{This fact can be confirmed by direct computation.
It can also be proven by use of the parity operation
defined in \secref{sec:NoMystery}. 
Let us tag one of the impurities, $\phi'$, 
$D_2$ will only contract to this impurity.
Whatever the outcome of $D_2 \Tr \phi' Z^p \phi Z^{J-p}$ will
be, it is again of the type~\eqref{symop} and thus has positive parity. 
Therefore
$D_2 \Tr \phi' Z^p \phi Z^{J-p}=
PD_2 \Tr \phi' Z^p \phi Z^{J-p}=
D_2P \Tr \phi' Z^p \phi Z^{J-p}=
D_2  \Tr Z^{J-p} \phi Z^p \phi'=
D_2  \Tr \phi Z^p \phi'\Tr Z^{J-p}$.}.
The operator $D_2\Op_p$ is 
again a sum of operators of the type~\eqref{symop}. Therefore, when evaluating
$D_2(D_2 \Op_p)$ we get the same contribution from the two possible
contractions of the $\check{\phi}$ in the leftmost $D_2$ with a $\phi$ in
$(D_2\Op_p)$. We can thus determine $D_2(D_2 \Op_p)$ by contracting 
$\check{\phi}$ 
in the leftmost $D_2$ with the $\phi$ in the already applied $D_2$ and 
multiplying the result by two.
 Next, the $\check{Z}$ of
the leftmost $D_2$ must be contracted with one of the
$Z$'s of $(D_2\Op_p)$. This can either be a $Z$ of the original
$\Op_p$ or the $Z$ appearing in the already applied $D_2$. These
two types of contractions are, up to an overall factor, \
described respectively by the
vertices~\eqref{Z2} and~\eqref{Noneloop}.

It is easy to write down the \emph{exact} expression for
$D_2\Op_p$. Let us define
\[\label{D2}
D_2\equiv N D_{2;0}+D_{2;+}+D_{2;-},
\]
where $D_{2;0}$ is trace conserving and $D_{2;+}$ and $D_{2;-}$
respectively increases and decreases the trace number by one. 
Then we find
\<\label{D20}
{D}_{2;0}\, \Op_p^{J_0;J_1,\ldots,J_k}\eq
-4\Big(\delta_{p\neq J_0}\Op^{J_0;J_1,\ldots,J_k}_{p+1} 
-(\delta_{p\neq J_0}+\delta_{p\neq 0})\, \Op^{J_0;J_1,\ldots,J_k}_p 
\nlnum 
\qquad+\delta_{p\neq 0}\Op^{J_0;J_1,\ldots,J_k}_{p-1} 
\, \Big)  ,
\nln
D_{2;+}\, \Op_p^{J_0;J_1,\ldots,J_k}\eq
4\sum_{J_{k+1}=1}^{p-1}\, 
\left( 
\Op^{J_0-J_{k+1};J_1,\ldots,J_{k+1}}_{p-J_{k+1}} -
\Op_{p-1-J_{k+1}}^{J_0-J_{k+1};J_1,\ldots,{J_{k+1}}} 
\right) 
\nl
-4\sum_{J_{k+1}=1}^{J_0-p-1}\,
\left( 
\Op^{J_0-J_{k+1};J_1,\ldots,J_{k+1}}_{p+1}
-\Op^{J_0-J_{k+1};J_1,\ldots,J_{k+1}}_p 
\right),
\nln
D_{2;-}\, \Op_p^{J_0;J_1,\ldots,J_k}\eq
4\sum_{i=1}^k J_i\, 
\left(
\Op^{J_0+J_i;J_1,\ldots,\makebox[0pt]{\,\,\,\,$\times$}J_{i},\ldots,J_k}_{J_i+p} 
-\Op^{J_0+J_i;J_1,\ldots,\makebox[0pt]{\,\,\,\,$\times$}J_{i},\ldots,J_k}_{J_i+p-1}
\right) 
\nl
-4\sum_{i=1}^k J_i\, \left( 
\Op^{J_0+J_i;J_1,\ldots,\makebox[0pt]{\,\,\,\,$\times$}J_{i},\ldots,J_k}_{p+1}
-\Op^{J_0+J_i;J_1,\ldots,\makebox[0pt]{\,\,\,\,$\times$}J_{i},\ldots,J_k}_p
\right). \nn
\>
In analogy with~\eqref{D2} one can write $\delta D_4$ as
\[\label{D4}
\delta D_4=N^2\, \delta D_{4;0}+N\,\delta D_{4;+}+N\,\delta D_{4;-}+\delta D_{4;++}+\delta D_{4;--}+\delta D_{4;+-},
\]
where $\delta D_{4;+-}$ stands for the genus one trace number conserving part.
For $\delta {D}_{4;0}$ we get (for $J_0>0$)
\[\label{DD04}
\delta {D}_{4;0}\, \Op_p^{J_0;J_1,\ldots,J_k}=
4(\delta_{p,0}+\delta_{p,J_0}-\delta_{p,1}-\delta_{p,J_0-1})
\lrbrk{\Op^{J_0;J_1,\ldots,J_k}_{1}-\Op^{J_0;J_1,\ldots,J_k}_{0}}.
\]
The remaining contributions to $\delta D_4$ likewise always produce
operators in the combination $\Op_1-\Op_0$. We defer the exact expressions
for $\delta D_{4;-}$, $\delta D_{4;+}$, $\delta D_{4;++}$ and 
$\delta D_{4;+-}$ to \appref{deltaD4}. The contribution 
$\delta D_{4;--}$ is easily seen to vanish.

\subsection{The planar limit}
\label{ssec:TwoImpPlanar}

In the basis of symmetric two-impurity single trace states of the
type~\eqref{symop} with $k=0$
the planar one-loop dilatation matrix takes the form (for $J\geq 1$)
\[\label{dilplanar1}
D_{2;0}=4\cdot\left(\begin{array}{cccccc}
+1&-1&  &&&    \\
-1&+2&-1&  &  &  \\
  &\ddots&\ddots& \ddots& &  \\
  &  &\ddots&\ddots&\ddots&\\
  &  &  &-1&+2&-1\\
  &  &  &  &-1&+1 \\
\end{array}\right).
\]
Notice the special form of the boundary contributions which follows
from the presence of the $\delta$-functions in equation~\eqref{D20}.
As exploited in reference~\cite{Beisert:2002tn} the matrix~\eqref{dilplanar1}
has the following exact eigenvectors  
\[
\Op_n^J=\frac{1}{J+1}\sum_{p=0}^J
\cos\left(\frac{\pi n(2p+1)}{J+1}\right) \Op_p^J.
\label{OfiniteJ}
\]
We emphasize that equation \eqref{OfiniteJ} thus gives for any value of 
$J$ our one-loop planar eigenoperators.
Note that $\Op^{J}_{n}=\Op^{J}_{-n}=-\Op^{J}_{J+1-n}$. 
Thus the set of independent modes 
can be chosen as given by the mode numbers $0\leq n\leq[J/2]$.
The inverse transform is 
\[
\Op_p^J=\Op_{n=0}^J+2\sum_{n=1}^{[J/2]}
\cos\left(\frac{\pi n(2p+1)}{J+1}\right) \Op_n^J.
\label{idct}
\]
The corresponding exact planar one-loop anomalous dimension was 
found in~\cite{Beisert:2002tn}. 
Moving on to the two-loop analysis
we infer from~\eqref{deltaD} and \eqref{D4} that the matrix $D_{4,0}$
can be expressed as the square of 
the one-loop matrix \eqref{dilplanar1}
plus contact-term contributions from \eqref{DD04}
\[\label{eq:dD4mat}
D_{4;0}=-\sfrac{1}{4}(D_{2;0})^2+
4\cdot\left(\begin{array}{cc|ccc|cc}
-1&+1&  &  &  &  &  \\
+1&-1&  &  &  &  &    \\\hline
  &  & 0&  &  &  & \\
  &  &\phantom{+1}&\ddots&\phantom{+1}&  & \\
  &  &  &  & 0&  &  \\\hline
  &  &  &  &  &-1&+1\\
  &  &  &  &  &+1&-1
\end{array}\right).
\]
We now face the problem that the
states \eqref{OfiniteJ} are no longer eigenstates of $D_{4;0}$, since
$D_{2;0}$ and $\delta D_{4;0}$ do not commute. 
We find
\[
\delta D_{4;0} \Op_n^J=
-\frac{256}{J+1}
\sin^2\frac{\pi n}{J+1} \cos\frac{\pi n}{J+1}
\sum_{m=1}^{[J/2]}
\sin^2\frac{\pi m}{J+1} \cos\frac{\pi m}{J+1}\,
\Op_m^J.
\]
However, we can treat $D_{4;0}$ as a perturbation and thus find
that the two-loop part of the planar anomalous dimension 
is the diagonal piece of $D_{4;0}$ proportional
to $\Op^J_n$.
Exploiting the relation~\eqref{deltaD}, 
for the operators whose one-loop form is
given by~\eqref{OfiniteJ} 
we obtain the following planar anomalous dimension
exact to two loops
\[\label{andim}
\Delta_n^J=J+2+
\frac{\gym^2 N}{\pi^2}\,\sin^2\frac{\pi n}{J+1}-
\frac{\gym^4 N^2}{\pi^4}\,\sin^4\frac{\pi n}{J+1}
\lrbrk{\frac{1}{4}+\frac{\cos^2\frac{\pi n}{J+1}}{J+1}}.
\]
This formula comprises the results 
$\Delta_{\mathcal{K}'},\Delta\indup{b}$ in 
\eqref{eq:singledim} as well as the planar piece in 
$\Delta\indup{single}$ in 
\eqref{eq:BianchiSingle},
\eqref{disexp} and
\eqref{planarJ5}.
In \eqref{finiteJ3} we present a conjecture for the planar three-loop
contribution to $\Delta_n^J$.
Furthermore, using standard perturbation theory we can also find the 
(planar) perturbative correction to the 
eigenstates \eqref{OfiniteJ}:
They involve the coupling constant dependent redefinition
\[\label{eq:twoimpredef}
\Op_n^J \rightarrow \Op_n^J- \frac{\gym^2 N}{\pi^2}\,\frac{1}{J+1}\,
\sum_{m =1 }^{[J/2]}
\delta_{m\neq n}
\frac{\sin^2\frac{\pi n}{J+1} \cos\frac{\pi n}{J+1}
\sin^2\frac{\pi m}{J+1} \cos\frac{\pi m}{J+1}}
{\sin^2\frac{\pi n}{J+1} - \sin^2\frac{\pi m}{J+1}}~
\Op_m^J.
\]
This $\gym$-dependent remixing is a complicating feature that 
we can expect at each further quantum loop order; remarkably,
it is absent in the small $J$ limit (i.e.~for $J=2$ (Konishi)
and $J=3$) and in the large $J$ (BMN) limit
as discussed in the next section.

\section{The BMN limit of two impurity operators}
\label{sec:BMNlimit}

In the BMN limit one is supposed to send $J$ to infinity and consider
only operators for which $D_0-J$ is finite~\cite{Berenstein:2002jq}. 
More precisely,
the BMN limit is defined as the double scaling 
limit~\cite{Kristjansen:2002bb,Constable:2002hw} 
\[
J\rightarrow \infty, \hspace{0.7cm} N\rightarrow \infty,
\hspace{1.0cm}\lambda'=\frac{\gym^2 N}{J^2}, \,\,\,\,\,
g_2=\frac{J^2}{N}\hspace{0.5cm} \mbox{fixed}. \label{BMNlim}
\]
In reference~\cite{Beisert:2002ff} it was shown how the one-loop 
dilatation operator
could be identified as a certain quantum mechanical Hamiltonian
in the BMN limit. Since the same quantum mechanical Hamiltonian turns out
to contain all information about the two-loop dilatation operator in the
BMN limit we shall briefly recall its derivation. 

\subsection{The one-loop Hamiltonian}
\label{ssec:BMNcontinuum}

With $J$ being very large
we can view $x\equiv\frac{p}{J}$ and $r_i\equiv\frac{J_i}{J}$ as continuum
variables and we replace the discrete set of states in equation~\eqref{symop}
by a set of continuum states
\[
\label{states}
\Op_p^{J_0;J_1,\ldots,J_k} \rightarrow |x;r_1, \ldots, r_k \rangle
=|r_0-x;r_1, \ldots, r_k \rangle,
\]
where
\[
x \in [0,r_0],\qquad
r_0,r_i\in [0,1] \qquad \mbox{and}\qquad r_0=1-(r_1+ \ldots +r_k),
\]
and where it is understood that 
$|x;r_1, \ldots, r_k \rangle = |x;r_{\pi(1)}, \ldots, r_{\pi(k)} \rangle$
with \mbox{$\pi \in \;$S$_k$} an arbitrary permutation of $k$ elements.

Defining
\[
g^2D_2=\frac{\lambda'}{4\pi^2}\,d_2,
\]
we can write $d_2=d_{2;0}+g_2\, d_{2;+}+g_2\, d_{2;-}$
and imposing the BMN limit~\eqref{BMNlim} we get
a continuum version of the equations~\eqref{D20}
\<
\label{haction}
d_{2;0}\, |x;r_1, \ldots, r_k \rangle \eq -\partial_x^2~
|x;r_1, \ldots, r_k \rangle  ,
\\\nn
d_{2;+}\, |x;r_1, \ldots, r_k \rangle \eq
\int_0^x dr_{k+1}\, ~\partial_x 
~|x-r_{k+1};r_1, \ldots, r_{k+1} \rangle
\nl
-\int_0^{r_0-x}dr_{k+1}\, 
~\partial_x 
~|x;r_1, \ldots, r_{k+1} \rangle ,
\nln
d_{2;-}\, |x;r_1, \ldots, r_k \rangle \eq
\sum_{i=1}^k ~r_i~\partial_x~
|x+r_i;r_1, \ldots, \makebox[0pt]{\,\,$\times$}r_{i},\ldots,r_k \rangle
\nl
-\sum_{i=1}^k ~r_i~\partial_x~
|x;r_1, \ldots, \makebox[0pt]{\,\,$\times$}r_{i},\ldots,r_k \rangle.
\nn
\>
The $(k+1)$-trace eigenstates at $g_2=0$ are (with $n$ integer)%
\footnote{Notice that as opposed to in reference~\cite{Beisert:2002ff} we use symmetrized
operators and therefore the cosine transform appears.}
\[
\label{fourier}
|n;r_1, \ldots, r_k \rangle = \frac{1}{r_0}\,
\int_0^{r_0} dx\, \cos\left(2\pi n x/r_0\right)\, 
|x;r_1, \ldots, r_k \rangle.
\]
This is of course in accordance with the nature of the exact eigenstates
at finite $J$, \emph{cf}.\ eqn.~\eqref{OfiniteJ}. 
The inverse transform of~\eqref{fourier}
reads
\[ \label{cosine}
|x;r_1, \ldots, r_k \rangle=
|0;r_1, \ldots, r_k \rangle+
2\sum_{m=1}^{\infty}\cos\left(2\pi m x/r_0\right)
|m;r_1, \ldots, r_k \rangle.
\]
In the basis~\eqref{fourier} the action of the operator $d_2$ reads
\<\label{h-on-eigen}
d_{2;0}\, |n;r_1,\ldots, r_k\rangle \eq
\bigbrk{\sfrac{2\pi n}{r_0}}^2\, |n;r_1,\ldots, r_k\rangle,
\\
d_{2;+}\,  |n;r_1, \ldots, r_k \rangle \eq
\frac{8}{r_0}\int_0^{r_0} dr_{k+1}\, \sum_{m=1}^{\infty}
\frac{\bigbrk{\frac{2\pi m}{r_0-r_{k+1}}}^2\sin^2\bigbrk{\pi n\sfrac{r_{k+1}}{r_0}}}
{\bigbrk{\frac{2\pi m}{r_0-r_{k+1}}}^2-\bigbrk{\frac{2\pi n}{r_0}}^2}
~|m;r_1, \ldots, r_{k+1} \rangle ,
\nln\nn
d_{2;-}\,  |n;r_1, \ldots, r_k \rangle \eq
8\sum_{i=1}^k \frac{r_i}{r_0} \sum_{m=1}^{\infty}
\frac{\bigbrk{\frac{2\pi m}{r_0+r_i}}^2 \sin^2\bigbrk{\pi m\sfrac{r_i}{r_0+r_i}}}
{\bigbrk{\frac{2\pi m}{r_0+r_i}}^2-\bigbrk{\frac{2\pi n}{r_0}}^2}
~|m;r_1, \ldots, \makebox[0pt]{\,\,$\times$}r_{i},\ldots,r_k \rangle .
\>
Now, the scene is set for determining the spectrum of the full one-loop
Hamiltonian order by order in $g_2$ by standard quantum mechanical
perturbation theory and in reference~\cite{Beisert:2002ff} we carried out
this program for single trace operators to three leading orders in $g_2$.
Note the great simplicity of the expressions eq.\eqref{h-on-eigen}
(no contact terms etc.), emerging directly from the study of the
dilatation operator. Rather than comparing specific consequences of
the above equations, and turning around the suggestion in 
\cite{Gross:2002mh,Gomis:2002wi,Gomis:2003kj}, 
we feel that it would be important to derive the above Hamiltonian
in the string field formulation of plane wave
strings 
\cite{Spradlin:2002ar,Klebanov:2002mp,Spradlin:2002rv,
Pankiewicz:2002gs,Pankiewicz:2002tg,
He:2002zu,Roiban:2002xr}
or the (interpolating) string bit formulation of 
\cite{Verlinde:2002ig,Vaman:2002ka,Pearson:2002zs}.

\subsection{Two loops}
\label{ssec:BMN2Loop}

Proceeding to two loops we define
\[
g^4D_4=\left(\frac{\lambda'}{4\pi^2}\right)^2 d_4,
\]
and we see from equation~\eqref{deltaD} that the only new element in
the analysis is the determination of the BMN limit of the operator
$\delta D_4$. Now, it follows from the equations in \appref{sec:deltaD4}, 
that acting with $\delta D_4$ on a state of 
the type~\eqref{symop} produces at the discrete level only states in the
combination $\Op_1-\Op_0$. Expressing this quantity in terms of the
exact eigenstates at $g_2=0$, \eqref{OfiniteJ}, it is easily seen that
in the BMN limit we have%
\footnote{Notice that $\Op_1-\Op_0$ is down
by a factor of $\frac{1}{J}$ compared to $\Op_p-\Op_{p-1}$ for general
$p$. In the continuum language this is reflected by the fact that
$\partial_x|x;r_1,\ldots,r_k\rangle|_{x=0}=0$ which follows from the
relation~\eqref{cosine}.}
\[
\Op_1^{J_0;J_1,\ldots,J_k}-\Op_0^{J_0;J_1,\ldots,J_k}
\sim \frac{1}{J_0^2}.
\]
Using this result a straightforward scaling analysis of the relations
in \appref{sec:deltaD4}
shows that the operator
$\delta D_4$ becomes irrelevant in the 
BMN limit. 
Thus, in this limit we have
\[
d_4=-\quarter (d_2)^2,
\]
and at any given order in the genus expansion, 
the eigenstates of the one-loop
Hamiltonian are also eigenstates of the two-loop Hamiltonian%
\footnote{This proves that the $\order{\lambda'}$ 
contribution to the three-point function
of symmetric-traceless operators $\Op^{\prime\prime\,J}_{(ij),n}$
in \cite{Beisert:2002bb} is indeed correct.
For the singlet or antisymmetric operators this remains
an open question.}. 
For the BMN limit of $D-J$ one gets up to and including two loops
\[\label{BMNdil}
D-J\longrightarrow 2+\left(\frac{\lambda'}{4\pi^2}\right)d_2-\frac{1}{4}
\left(\frac{\lambda'}{4\pi^2}\right)^2(d_2)^2.
\]
Writing the BMN double expansion of the conformal dimension of an operator
as
\[
\Delta(\lambda',g_2)-\Delta_0=\sum_{k=1}^{\infty}
\left(\frac{\lambda'}{4\pi^2}\right)^k 
\Delta_k(g_2),
\]
with $\Delta_0$ the tree-level conformal dimension, what we learned
above can be expressed as
\[
\Delta_2(g_2)=-\quarter \bigbrk{\Delta_1(g_2)}^2,
\qquad \forall\,\,\, g_2.
\]
\subsection{Degeneracies}

{}From equation~\eqref{h-on-eigen} it follows that at the planar, one-loop
level we have a degeneracy between states $|n;r_1,\ldots,r_k\rangle$ and
$|n';r_1',\ldots,r_k'\rangle$ for which $\frac{n}{r_0}=\frac{n'}{r_0'}$.
Focusing on the simplest case a single trace state $|n\rangle$ is degenerate
with double trace states of the type $|m;r\rangle$ for which 
$n=\frac{m}{1-r}$, with triple trace states $|k;r_1,r_2\rangle$ 
for which $n=\frac{k}{1-r_1-r_2}$ and so on \cite{Constable:2002vq}. 
Notice that degeneracy is excluded for the special case $|n\rangle=|1\rangle$. 
Determining the energy shift of the state $|n\rangle$ with $n>1$ to genus one
by treating the term $g_2(d_{2,+}+d_{2,-})$ as a perturbation thus 
in principle requires the use of degenerate perturbation theory. It can be
shown that at order $g_2$ matrix elements between single trace states
and double trace states vanish~\cite{Beisert:2002bb,Constable:2002vq}.
In contrast, matrix elements between 
states $|n\rangle$ and $|k;r_1,r_2\rangle$ 
for $n=\frac{k}{1-r_1-r_2}$ are finite~\cite{Beisert:2002ff}
at order $g_2^2$. 
So far these degeneracy effects have not
been rigorously taken into account due to the technical obstacles 
caused by the fact that a single trace state is degenerate with a
continuum of triple trace states%
\footnote{One way to treat 
the degeneracy might be 
to consider the finite $J$ operators discussed 
in \secref{sec:twoimp} as  regularized BMN operators 
and then ensure that all
relevant degeneracies are
resolved before taking $J$ to infinity.
For single-trace operators this would require working with $n$ being a
divisor of $J+1$.}.
Bearing in mind the implications of the finite-$J$
degeneracies, \emph{cf}.\ \secref{ssec:degen}, it is most 
desirable to stringently carry through analysis in 
the large-$J$ case as well.

\subsection{All loop conjecture}
\label{ssec:BMNallloop}

Assuming that the effective vertex idea works at any loop order one would
expect that $d_2$ encodes all information about the BMN limit of the
dilatation operator in the basis~\eqref{symop}. In the BMN limit we
single out those terms of the dilatation operator which produce the
maximum number of $J$ factors. These are the terms with the maximum
number of $\check{Z}$ constituents. Therefore, only terms of the
dilatation operator which have only one impurity will survive the limit. 
Any such term will appear in some higher power of $d_2$ and
it has to appear with a weight which ensures
that the planar BMN limit 
\cite{Berenstein:2002jq,Gross:2002su,Santambrogio:2002sb}
\[\label{eq:BMNsz}
\Delta\longrightarrow J+2\,\sqrt{1+\vphantom{\dot W}\lambda'\, n^2},
\]
is conserved.
Thus, it is tempting to conjecture that the all loop version of the
formula~\eqref{BMNdil}
reads
\[\label{eq:BMNallloop}
D-D_0\longrightarrow 2\,\sqrt{1+
\left(\frac{\lambda'}{4\pi^2}\right)d_2}.
\]
We now construct a possible all-loop expression for the
one impurity part of the full dilatation generator which
manifestly gives rise to the above
formula in the case of two impurity operators. Iterating the
argument given in the beginning of \secref{ssec:bmn} we have
the following simplification when $(D_2)^l$ acts on two-impurity
operators
\[\label{eq:D2l}
(D_2)^l \sim
2^{l-1}
\phi^{(a_0)} D_{2}^{a_0a_1}
D_{2}^{a_1a_2}\ldots
D_{2}^{a_{l-1}a_{l}}
\check\phi^{(a_l)},
\]
where
\[
D_{2}^{ab}=
-2\normord{\Tr\comm{Z}{T^a}\comm{\check Z}{T^b}}.
\]
Thus, a possible form for the one-impurity part of the full $l$-loop
dilatation generator is
\[\label{eq:single-impurity}
D_{2l}=
(-2)^{1-l}\frac{(2l-2)!}{(l-1)!l!}\, \phi^{(a_0)} D_{2}^{a_0a_1}\,
D_{2}^{a_1a_2}\ldots
D_{2}^{a_{l-1}a_{l}}
\check\phi^{(a_l)}
+\ldots,
\]
up to terms irrelevant in the BMN limit.
There might, however, be other combinations which also give
rise to \eqref{eq:BMNallloop}.
At the planar level \eqref{eq:single-impurity}
is equivalent to
the conjecture of \cite{Berenstein:2002jq,Gross:2002su} and
the result of \cite{Santambrogio:2002sb}.
However, our possible formula would apply to arbitrary genus.
If it was correct, one would be able to
perform all-loop computations not only for
two-impurity BMN operators, but also for more
than two impurities.
There, we do not find a reason why $D_{2l}$ should be
proportional to $(D_2)^l$ and an individual 
genus expansion would be required each loop level.

The formula \eqref{eq:single-impurity}, if true, 
would also have implications for 
$\superN=4$ SYM beyond the BMN limit.
It successfully predicts the dilatation generator at
two-loops completely:
$D_4$ consists of three parts which can be labelled
as $(1,1)$, $(2,1)$ and $(1,2)$, the numbers representing the
number of interacting background and impurity fields $Z,\phi$.
The parts with one impurity are predicted directly,
the part $(1,2)$ is related to $(2,1)$ by symmetry.
 At the three-loop level all terms
except the terms with two impurities and two background
fields, $(2,2)$, can be predicted. As this represents a three-loop
interaction between four fields, the corresponding
Feynman-diagrams do not have bulk loops.
The complexity of such diagrams is comparatively low
and a direct computation of coefficients may still be feasible.

\section{Symmetry enhancement at the planar level and integrability}
\label{sec:NoMystery}

Let us define a parity operation $P$ acting on traces of group 
generators $T^a$ by inverting the order of generators within a trace
or, alternatively, by transposing all generators $T^a$. 
For example
\[\label{eq:parity}
P \Tr Z^3\phi^2Z\phi=\Tr \phi Z \phi^2 Z^3=\Tr Z^3 \phi Z \phi^2. 
\]
It is easily seen that the dilatation generators 
$D_0$, $D_2$, $D_4$ commute with the parity operation $PD_{2l}P=D_{2l}$.
In fact, parity is related to complex conjugation 
on group generators, $P T^a=(T^a)^\trans=(T^a)^\conj$.
As the dilatation generator $D$ is
real it conserves parity in general
\[\label{eq:parityconserved}
PD=DP.
\]
This implies that there is no mixing between operators of
different parity%
\footnote{The reason why the separation of parities 
has not been an issue so far is that
at least three impurities are required
to allow both parities within the same
$\grSO(6)$ representation.},
which was also noticed in \cite{D'Hoker:2003vf}.
This parity operation is specific to the $\grSU(N)$ series.
In the $\grSO(N)$ and $\grSp(N)$ series the parity of
a trace of $n$ generators $T^a$ is given by $(-1)^n$
due to the identity $P T^a=(T^a)^\trans=-T^a$
\footnote{A consequence of this is that many of the operators
discussed so far are incompatible with the $\grSO(N)$ and $\grSp(N)$ series.}.

As parity is conserved one should a priori expect that 
the spectra of positive and negative parity operators are not related. 
An example of a system with conserved parity 
is the infinite potential well.
There, the even and odd modes have distinct spectra which are
not related to each other in any obvious way. 
This is what happens in $\superN=4$ SYM in general.
If we however consider the strict planar limit, $N=\infty$,
we find that the two sectors are related. 
We observe that, whenever parity even and odd
states have equal quantum numbers (except parity), they form
pairs of operators with \emph{degenerate} scaling dimension. 
As a systematic degeneracy almost inevitably indicates a \emph{symmetry}
there seems to be a symmetry enhancement in the planar limit of
$\superN=4$ SYM. 

We start by discussing the most simple example of such a planar parity pair 
and afterwards we explain the degeneracy at the one-loop level
by means of an integrable spin chain \cite{Minahan:2002ve}.

\subsection{Planar parity pairs}
\label{ssec:planardegen}

There are three unprotected operators in the 
representation $[3,1,3]$ with $\Delta_0=7$.
One has negative parity
\[\label{eq:pairminus}
\Op_-=\Tr \comm{\phi}{Z}\comm{\phi}{Z}\comm{\phi}{Z}Z,
\]
and two have positive parity
\[\label{eq:pairplus}
\Op_+=\left(\begin{array}{c}
2\Tr Z^4\phi^3+2\Tr Z^2\phi Z^2\phi^2+2\Tr Z^2\phi Z\phi Z\phi-3 \Tr Z^3\phi\acomm{\phi}{Z}\phi\\
\Tr Z\phi \Tr Z^2\comm{\phi}{Z}\phi-\Tr Z^2 \Tr Z\comm{\phi}{Z}\phi^2
\end{array}\right).
\]
The scaling dimension for the first is readily obtained 
\[\label{eq:pairdim}
\Delta_-=7+\frac{5\gym^2N}{8\pi^2}-\frac{15\gym^4 N^2}{128\pi^4}.
\]
The dilatation generator acts on the other two as
\[
D_2=N\left(\begin{array}{cc}
10&\sfrac{8}{N}\\
\sfrac{20}{N}&8
\end{array}\right),\qquad
D_4=N^2\left(\begin{array}{cc}
-30-\sfrac{40}{N^2}&-\sfrac{40}{N}\\
-\sfrac{140}{N}&-24-\sfrac{64}{N^2}
\end{array}\right),
\]
corresponding to the scaling dimensions 
(the one loop part of which has been found in \cite{Ryzhov:2001bp})
\[
\Delta_+=7+\frac{\gym^2N}{16\pi^2}\lrbrk{9\pm\sqrt{1+\sfrac{160}{N^2}}}
-\frac{\gym^4 N^2}{256\pi^4}\lrbrk{27+\frac{52}{N^2}\pm\frac{3+\sfrac{948}{N^2}}{\sqrt{1+\sfrac{160}{N^2}}}}.
\]
Intriguingly we find that the
scaling dimension of the two single-trace operators,
$\Op_-$ and $\Op_{+,1}$ approach each other
for $N\to\infty$ at one-loop as well as at two-loop order.
We hence find a pair of operators with opposite parity
which have degenerate scaling dimensions.
This is actually not an exception, it is rather the \emph{rule}. 
Among the operators in the representations
$[3,\Delta_0-6,3]$ with $6\leq \Delta_0\leq 10$ 
we find $8$ pairs of operators and only $3$ unpaired ones,
\emph{cf}.\ \tabref{tab:onetrace}, \secref{sec:twoimp} and 
\appref{sec:anodim}. 
And they are not the only examples, we find them
in the representations
($\Delta_0=9$, $[4,1,4]$),
($\Delta_0=10$, $[4,2,4]$),
($\Delta_0=6$, $[0,2,0]$) and
($\Delta_0=6$, $[1,0,1]$).
In fact, every representation that admits both parities
seems to have such pairs of operators. 
\begin{table}\centering
\begin{tabular}[t]{|c|c|c||cc|}\hline
dim & $\grSO(6)$ &$\grSU(2)$ & $+$ & $-$
\\\hline\hline
4&[2,0,2]&$\rep{1}$&1&-
\\\hline
5&[2,1,2]&$\rep{2}$&1&-
\\\hline
6&[2,2,2]&$\rep{3}$&2&-\\
 &[3,0,3]&$\rep{1}$&-&1
\\\hline
7&[2,3,2]&$\rep{4}$&2&-\\
 &[3,1,3]&$\rep{2}$&1&1
\\\hline
8&[2,4,2]&$\rep{5}$&3&-\\
 &[3,2,3]&$\rep{3}$&1&2\\
 &[4,0,4]&$\rep{1}$&3&-
\\\hline
9&[2,5,2]&$\rep{6}$&3&-\\
 &[3,3,3]&$\rep{4}$&3&3\\
 &[4,1,4]&$\rep{2}$&3&1
\\\hline
\end{tabular}\qquad%
\begin{tabular}[t]{|c|c|c||cc|}\hline
dim & $\grSO(6)$ &$\grSU(2)$ & $+$ & $-$
\\\hline\hline
10&[2,6,2]&$\rep{7}$&4&-\\
  &[3,4,3]&$\rep{5}$&3&4\\
  &[4,2,4]&$\rep{3}$&8&2\\
  &[5,0,5]&$\rep{1}$&-&4
\\\hline
11&[2,7,2]&$\rep{8}$&4&-\\
  &[3,5,3]&$\rep{6}$&5&5\\
  &[4,3,4]&$\rep{4}$&10&5\\
  &[5,1,5]&$\rep{2}$&6&6
\\\hline
12&[2,8,2]&$\rep{9}$&5&-\\
  &[3,6,3]&$\rep{7}$&6&7\\
  &[4,4,4]&$\rep{5}$&17&7\\
  &[5,2,5]&$\rep{3}$&9&14\\
  &[6,0,6]&$\rep{1}$&12&2
\\\hline
\end{tabular}

\caption{
Single-trace pure scalar operators of 
dimension $\Delta_0\leq 12$.
The operators are distinguished by their 
representation and parity.
The protected operators of 
representation $[0,\Delta_0,0]$ were omitted.
The operators correspond to states 
of a $\grSU(2)$ spin chain of length
$L=\Delta_0$.}
\label{tab:onetrace}
\end{table}

This phenomenon is a signal of an enhanced symmetry in the 
planar limit of the $\grSU(N)$ series%
\footnote{The groups $\grSO(N)$, $\grSp(N)$, although admitting a planar limit, 
do not show this behavior: One of the parity partners is always absent.}.
It cannot be supersymmetry (this time) for a simple reason:
Supersymmetry is independent of the choice of gauge group whereas 
at finite $N$ the operators $\Op_+$ mix and have modified
anomalous dimensions. 
Furthermore, for gauge groups other than $\grSU(N)$ the pairs 
cannot even exist.
Therefore the operators $\Op_{+,1}$ and
$\Op_-$ cannot be part of the same supermultiplet.
As the multiplets of this symmetry seem to be either singlets
or doublets the symmetry is most likely abelian.
In the next subsection we proceed by proving the degeneracy
at the one-loop level. 

\subsection{A conserved charge}
\label{ssec:bonussym}

In order to prove the existence of an abelian symmetry 
in the planar sector we will find a generator $U$ and show
that it commutes with the generator of dilatations $D$.
Furthermore, we demand that $U$ anticommutes with the
parity operation $P$ so that it may relate operators of opposite parities. 
\[
\comm{D}{U}=\acomm{P}{U}=0.
\]
We will restrict ourselves to the one-loop level, 
where we can allow all six scalar fields;
mixing with other fields is a higher-order effect.
The degeneracy at the two-loop level will be investigated
in the next section.

\begin{figure}\centering
$\displaystyle
\parbox{1.5cm}{\centering\includegraphics{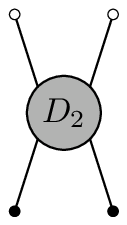}}
\quad=
\quad2\parbox{1.5cm}{\centering\includegraphics{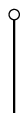}}
\quad-
\quad2\parbox{1.5cm}{\centering\includegraphics{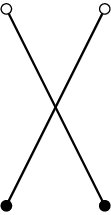}}
\quad+\quad
\parbox{1.5cm}{\centering\includegraphics{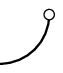}}
$
\caption{Graphical representation of the planar one-loop interaction $D_2$.}
\label{fig:oneloopplanar}
\end{figure}

We represent a single-trace operator by a cyclic
$\grSO(6)$-vector spin chain of length $L=\Delta_0$ 
with zero (angular) momentum as suggested by Minahan and Zarembo 
\cite{Minahan:2002ve}.
The one-loop dilatation operator is then given by 
\[
D_2=N \sum_{k=1}^L D_{2,k(k+1)},
\]
where $D_{2,k(k+1)}$ is a local interaction 
linking sites $k$ and $k+1$ (the sites are periodically identified).
The local interaction $D_{2,k(k+1)}$ is given by \cite{Minahan:2002ve}
\[\label{eq:D2spin}
D_{2,k(k+1)}=2I_{k(k+1)}-2P_{k(k+1)}+K_{k(k+1)}.
\]
Here, $I_{k(k+1)}$ is the identity,
$P_{k(k+1)}$ exchanges sites $k,k+1$ and $K_{k(k+1)}$ is a
$\grSO(6)$ trace over sites $k,k+1$. 
Graphically this may be represented as in \figref{fig:oneloopplanar}.

\begin{figure}\centering
$\displaystyle
\parbox{2.0cm}{\centering\includegraphics{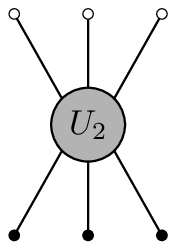}}
=
\parbox{2.0cm}{\centering\includegraphics{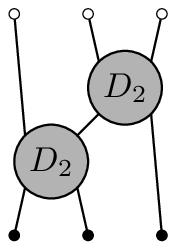}}
-
\parbox{2.0cm}{\centering\includegraphics{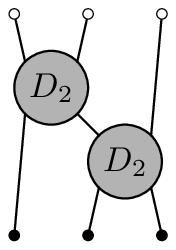}}
$
\caption{Graphical representation of the generator $U_2$.}
\label{fig:bonussym}
\end{figure}

We introduce a generator 
\[\label{eq:U2}
U_2=\sum_{k=1}^L U_{2,k(k+1)(k+2)},\qquad
U_{2,k(k+1)(k+2)}=\comm{D_{2,(k+1)(k+2)}}{D_{2,k(k+1)}}.
\]
Graphically it may be represented as in \figref{fig:bonussym}.
This generator is easily seen to anticommute with parity
(it has negative mirror symmetry w.r.t.\ the 
vertical axis, see \figref{fig:bonussym}). 
A straightforward but tedious calculation 
shows that $U_2$ indeed commutes with $D_2$.
We will explain this fact in terms of integrability in 
the next subsection,
\emph{cf.}~\eqref{eq:superduper}.

We note that $U_2$ interchanges 
$\Op_-$ of \eqref{eq:pairminus} with 
$\Op_{+,1}$ of \eqref{eq:pairplus},
\[\label{eq:U2action}
U_2\Op_{+,1}=-60 \Op_-,\qquad
U_2\Op_{-}=+4\Op_{+,1},
\]
so the generator $U_2$ is 
indeed responsible for the degeneracy of their scaling dimensions.
The same is true for all the other pairs of operators we observed.
The unpaired operators, for example \eqref{eq:singleops}, are
annihilated by $U_2$. 
As an aside we note that the eigenoperators are 
$\sfrac{1}{\sqrt{60}}\Op_{+,1} \pm \sfrac{i}{2}\Op_-$
\footnote{This could also be inferred by demanding that
the eigenoperators be orthonormal and related by $P$.}.
The corresponding eigenvalues of $U_2$ are 
$\pm 4i\sqrt{15}$ in this case.

Thus we have proven the existence of an additional abelian
symmetry in the planar sector at the one-loop level. 
The symmetry generated by $U_2$, however, cannot be
compact:
The eigenvalues of $U_2$ are not integer multiples of a common number. 
Nevertheless, one is led to believe that there exists a
$\grSO(2)$ symmetry (not generated by $U_2$)
which has uncharged singlets and charged doublets. 
Together with the parity operation $P$ it would form the
group O$(2)$. In this scenario, when $\frac{1}{N}$ corrections are included,
the group O$(2)$ breaks to the parity $\Integers_2$.
It would be very desirable to understand this degeneracy/symmetry 
better, in terms of $\superN=4$ SYM as well as in terms
of the AdS/CFT correspondence. 
Another peculiarity of the planar sector 
is a recently found degeneracy in four-point 
functions \cite{Arutyunov:2003ae}.
There it was observed that for $N\to\infty$ a four-point function
at one-loop could be described by a single function, although
by superconformal symmetry, two functions would be allowed.
Four-point functions are related to three-point functions and
anomalous dimensions by means of the operator product expansion. 
Therefore these issues might be related in some way.
It would also be of interest to study whether the
degeneracy can be observed as a symmetry of planar
$n$-point functions.

\subsection{Higher charges of the spin chain}
\label{ssec:highercharges}

A much more enlightening way to prove $\comm{D_2}{U_2}=0$
is to make use of integrability of the spin chain. 
In \cite{Minahan:2002ve} it was shown that the
interaction \eqref{eq:D2spin} has just the
right relative coefficients to exhibit integrability.
The corresponding $R$ matrix is
\[
R_{0k}(u)=P_{0k}\bigbrk{\brk{1-\sfrac{3}{2}u+\half u^2}I_{0k}
+u\brk{1-\sfrac{3}{2}u} \half D_{2,0k}
+\half u^2 (\half D_{2,0k})^2},
\]
where $D_{2,0k}$ is the local dilatation generator
\eqref{eq:D2spin} acting on an auxiliary site $0$, 
see \cite{Minahan:2002ve} for details. 
This $R$ matrix satisfies the Yang-Baxter equation for the $\grSO(6)$ case
\cite{Reshetikhin:1983vw,Reshetikhin:1985vd}
\[\label{eq:YB}
R_{12}(u)R_{13}(u+v)R_{23}(v)
=
R_{23}(v)R_{13}(u+v)R_{12}(u),
\]
and thus gives rise to an integrable spin chain.
Integrability predicts the existence of a tower
of commuting charges $t_n$. The zeroth charge is 
the cyclic shift, it equals 
the identity in the zero (angular) momentum sector. 
The first charge is identified with the dilatation generator
\[\label{eq:firstcharge}
t_1=\sfrac{1}{2N} D_2-\sfrac{3}{2}L,
\]
up to a constant. 
Using the expressions in \cite{Minahan:2002ve} we find for the
second charge
\[\label{eq:secondcharge}
t_2=-\sfrac{1}{8}U_2+\sfrac{1}{2}(t_1)^2-\sfrac{5}{8}L.
\]
{}From the fact that $t_1$ and $t_2$ commute we easily derive 
\[\label{eq:superduper}
\comm{D_2}{U_2}=0.
\]

We have seen that the second charge of the integrable spin chain
has important consequences. It is thus natural to 
investigate the third charge $t_3$. Up to polynomials
in $t_1$, $t_2$ and $L$ we find some generator $Q_{3,2}$ which
commutes with $D_2$, $U_2$ and parity $P$.
In contrast to $U_2$ we find that $Q_{3,2}$ does not pair up
operators, it simply assigns a number (charge) to each operator.
This is in fact what is to be expected. 
The reason why $U_2$ was interesting is that it 
anticommutes with $P$ while $D_2$ commutes thus giving rise to pairs. 
The next charge, $t_4$, will again give rise to some new
generator, $Q_{4,2}$, that anticommutes with parity. This generator will 
relate the same pairs, only with different coefficients (charges). 
Due to \eqref{eq:firstcharge} we know that the spectrum 
of $t_1$ is related to the spectrum of one-loop 
anomalous dimensions. A natural question to ask is whether
the spectra of the higher charges $Q_{2,2}=U_2$, $Q_{3,2}$, \ldots, 
have a physical meaning in the gauge theory. Except for special classes of operators,
the two-impurity operators for example, we find no obvious relation
between the spectra of $D_2$ and $Q_{3,2}$. 
Thus, there might be some non-trivial
information contained in the higher charges. 
Clearly, the deep question is \emph{why} integrability 
emerges from the planar gauge theory.

\section{Integrability at higher loops}
\label{sec:highergrounds}

In the previous section we have found planar parity pairs and
justified their existence using integrability at the one-loop level. 
At the two-loop level integrability has not been established yet.
An obstacle in doing so is that two-loop interactions are
interactions of next-to-nearest neighbors, whereas 
an integrable spin chain usually involves nearest neighbor interactions only.
Although some next-to-nearest neighbor interactions are included
in the tower of higher charges, these cannot be related to the
two-loop dilatation generator, because $D_2$ and $D_4$ do not commute 
while the spin chain charges do.
In order to construct an integrable spin chain with non-nearest
neighbor interactions we cannot make direct use of the
R-matrix formalism and the Yang-Baxter equation.
However, we may use our discovery of degenerate pairs in the previous section 
as the starting point.
Integrability at one-loop gives rise to a conserved charge $U_2$.
The charge anticommutes with parity $\acomm{P}{U_2}=0$
and thus pairs up operators.
By promoting $D_2$ and $U_2$ to their full counterparts
$D$ and $U$ we may generalize the integrable 
spin chain to higher loops or non-nearest neighbor interactions. 
The observed degeneracy at the two-loop level 
is a clear indication that a conserved $U$ exists up to two-loops.

By going to the two-loop level we again restrict ourselves to the operators
described in \secref{ssec:purescalars}. Effectively this means we 
consider a XXX$_{1/2}$ ($\grSU(2)$ spin $\half$) spin chain 
which does not have the trace term.
We can now represent all interactions by means of permutations
of adjacent sites. A generic term will be written as
\[\{n_1,n_2,\ldots\}=
\sum_{k=1}^L P_{k+n_1,k+n_1+1}
P_{k+n_2,k+n_2+1}\ldots\]
In this notation the one-loop dilatation generator 
(we will drop factors of $N$ in this section) and the 
charge $U_2$ are given by 
\[
D_{2}=2\bigbrk{\{\}-\{0\}},\qquad
U_2=4\bigbrk{\{1,0\}-\{0,1\}}.
\]
We start by showing that the degeneracy of parity pairs
holds at two-loop. {}From the assumption that 
planar $\superN=4$ SYM is integrable at three-loops
we derive the corresponding dilatation generator. 
Then we investigate the constraints on the four-loop dilatation 
operator and conclude
with an analysis of a tower of special three-impurity operators. 

\subsection{Two-loops}
\label{ssec:int2loop}

At the two-loop level we have found that the dilatation generator 
$D$ receives the correction
\[
D_4=2\bigbrk{-4\{\}+6\{0\}-\lrbrk{\{0,1\}+\{1,0\}}},
\]
and we expect that also $U$
receives some correction $U_4$. 
For the degeneracy to hold at the two-loop level 
we have to find a $U_4$ such that 
\[\label{eq:Uconserved4}
\comm{D_4}{U_2}+\comm{D_2}{U_4}=0, \hspace{0.5cm} 
\mbox{and}\hspace{0.5cm} \{P,U_4\}=0.\]
We find that this is satisfied by
\[
U_4=8\bigbrk{\{2,1,0\}-\{0,1,2\}}.
\]
This proves the degeneracy of anomalous dimensions 
at two-loops.
In performing the commutator we were required to make use of
the impossibility of three antisymmetric spins in $\grSU(2)$.
This indicates that for spin chains with next-to-nearest neighbor
interactions, integrability requires that 
for each site there can only be two states,
i.e.\ a $\grSU(2)$ spin chain.
This, however, does not exclude that integrability
extends to all six scalars of SYM. When the other scalars
are included, we have to consider fermions and derivative
insertions as well. This might be, roughly speaking,
a $\grSU(2,2|4)$ spin chain.
The interactions involving all
relevant fields might then be integrable.
At least for the superpartners of the
operators under consideration this must indeed be
the case due to supersymmetry.

In addition we have found the extension 
of the third charge $Q_{3}$ to two-loops,
see \appref{sec:highercharges}. 
This suggests that also
the higher charges will generalize to at
least two-loops.
If this is indeed so it justifies our claim that 
planar $\superN=4$ SYM at two-loops is integrable.
It is thus natural to conjecture that 
integrability holds at all-loops. 
This would mean there exist 
infinitely many commuting charges $Q_k$ with
loop expansion
\[
Q_k=\sum_{l=0}^\infty \left(\frac{\gym^2}{16 \pi^2} \right)^l
Q_{k,2l},
\]
where $Q_1=D$ and $Q_2=U$.
Considering the structure of 
the first few known terms
we assume that $Q_{k,2l}$ is composed of
up to $k+l-1$ permutations involving up to
$k+l$ adjacent sites. 
For $k$ odd the interactions are
symmetric and parity-conserving,
and for $k$ even they are 
antisymmetric and parity-inverting. 
Note that in the latter case this implies
a vanishing constant piece.

\subsection{Three-loops}
\label{ssec:int3loop}

In this subsection we investigate the constraints on the 
planar dilatation generator at three-loops due to 
integrability. This requirement together with the
correct behavior in the BMN limit, \eqref{eq:BMNsz},
fixes $D_6$ completely. 
This is not supposed to imply that the $D_6$ we derive 
is the correct planar dilatation generator of $\superN=4$ SYM. 
It may well be that the degeneracy of pairs is broken at
three-loops.

As building blocks for the three-loop interaction we allow
symmetric (with respect to the order of permutations 
$\{n_1,\ldots,n_k\}\mapsto\{n_k,\ldots,n_1\}$ to
ensure a real spectrum) and parity-conserving 
(positive symmetry under $\{n_1,n_2,\ldots\}\mapsto\{-n_1,-n_2,\ldots\}$)
structures.
These should act on only four adjacent sites and have no more than
three permutations. This follows from the general structure of a
connected three-loop vertex. 
We find exactly six structures that satisfy these constraints.
Using the planar BMN limit \eqref{eq:BMNsz}
we can fix the coefficients of four of these structures. 
The requirement of pairing for three-impurity operators at dimension 
$7$ and $8$,
see \secref{ssec:planardegen} and
\eqref{eq:dim8pair} fixes the remaining two. We find
\[
D_6=4\bigbrk{
15\{\}
-26\{0\}
+6\lrbrk{\{0,1\}+\{1,0\}}
+\{0,2\}
-\lrbrk{\{0,1,2\}+\{2,1,0\}}
}.
\]
The associated correction to $U$ 
can be found in much the same way 
analyzing all possible antisymmetric, parity-inverting structures
and requiring that $D$ and $U$ commute.
This fixes $U_6$ up to a contribution proportional 
to the fourth charge at one-loop $Q_{4,2}$. 
Acting on the Konishi descendant $\Tr\comm{\phi}{Z}\comm{\phi}{Z}$ we 
find the three-loop planar contribution to its scaling dimension
\[\label{eq:kon3}
\Delta_{\mathcal{K}'}
=4+\frac{3 \gym^2 N}{4\pi^2}
-\frac{3 \gym^4 N^2}{16\pi^4}
+\frac{21 \gym^6 N^3}{256\pi^6},
\]
subject to the assumption made at the beginning of this subsection.
It is also 
worth noting the contribution to a general two-impurity operator.
Up to a piece proportional to $(D_2)^3$ we find some contribution in 
the upper-left $3\times 3$ block (and equivalently in the
lower-right $3\times 3$ block)
\[
D_{6}=\sfrac{1}{8}(D_2)^3 
+4
\cdot\left(\begin{array}{ccc|cc}
+9&-10&+1&  &  \\
-10&+10&\phantom{+40}  &  &  \\
 +1&   &-1&  &  \\\hline
   &   &  & 0& \\
   &   &  &\phantom{+40}&\ddots  
\end{array}\right),
\]
in analogy to \eqref{eq:dD4mat}.
Considering now both $D_4$ and $D_6$ as perturbations of $D_2$ 
(\emph{cf}.\ \secref{ssec:TwoImpPlanar}) we can obtain 
the planar three-loop contribution to the
anomalous dimension of the operators~\eqref{OfiniteJ}. 
We get one
contribution
from the diagonal part of $D_6$ and one contribution from the
off-diagonal part of $D_4$ as encoded in the usual formula for second
order non-degenerate perturbation theory. The final result reads 
\[
\label{finiteJ3} \delta\Delta_n^J=\frac{\gym^6 N^3}{\pi^6} 
\sin^6\frac{\pi n}{J+1} 
\left(\frac{1}{8}+\frac{\cos^2\frac{\pi n}{J+1}}{4(J+1)^2}
\left(3J+2(J+6)\cos^2\frac{\pi n}{J+1}\right)\right),
\]
and it produces the correct result for all values of $J,n$,
even for the Konishi \eqref{eq:kon3} at $J=2$, $n=1$.

Using the planar three-loop dilatation generator
it might be possible derive the non-planar version
by analyzing all possible diagrams and fitting their
coefficients to non-renormalization theorems and the
planar version in analogy to \secref{sec:twoloop}.

\subsection{Four-loops}
\label{ssec:int4loop}

Having convinced ourselves of the usefulness of the constraints
from the pairing and the BMN limit we cheerfully proceed to 
four-loops, just to find that all coefficients but one are fixed. 
Unfortunately, the left-over coefficient does influence 
most four-loop anomalous dimensions, in particular the one of the Konishi
descendant $\mathcal{K}'$ 
\footnote{It is not even clear how to treat $\mathcal{K}'$, because the
interaction is longer than the spin chain in this case.}.
The consideration of higher charges might fix this remaining parameter.
Nevertheless, we will see below that we are able to 
find special operators for which the
four-loop
scaling dimensions turn out to be independent
of the so far undetermined parameter. 

At four-loops we find
in total twelve independent structures.
Five coefficients can be fixed using the planar BMN limit 
\eqref{eq:BMNsz}. Further five coefficients
are fixed by demanding degeneracy of pairs.
An expression for $D_8$ involving the remaining
two coefficients $\alpha,\beta$ can be found in \appref{sec:highercharges}.
Evaluating some eigenvalues of $D$ up to four-loops
one observes that they do not depend on the parameter $\beta$. 
This behavior is however expected. At four loops we have the
freedom to rotate the space of operators with the orthogonal 
transformation generated by the antisymmetric 
generator $\comm{D_2}{D_4}$. This gives rise
to the following similarity transformation 
\[D'=(1+\beta' \gnorm^6\comm{D_2}{D_4})^{-1}
D\,
(1+\beta' \gnorm^6\comm{D_2}{D_4}).\]
The first term in $D'$ due to the transformation is
\[
\beta'\gnorm^8\bigcomm{D_2}{\comm{D_2}{D_4}},
\]
and this is proportional to what multiplies $\beta$ in $D_8$. 
Now, as $D$ and $D'$ are related by a similarity transformation,
their eigenvalues are equal and $\beta$ only affects the
eigenvectors, but not the eigenvalues.

\subsection{A tower of three impurity operators}
\label{ssec:3tower}

For $\Delta_0$ odd we observe that there are as 
many even operators in the 
representation $[3,\Delta_0-6,3]$ as there are odd ones.
For a complete table of spin chain states, up to dimension 12,
see \tabref{tab:onetrace}. 
If $\Delta_0$ is even the same is true except for 
one additional operator with negative parity. 
This additional operator is 
\[\label{eq:special3op}
\Op=\sum_{k=1}^{\Delta_0-4}(-1)^k\Tr \phi Z^k\phi Z^{\Delta_0-3-k}\phi,\]
which is an exact planar eigenoperator of $D_2$ with eigenvalue $12N$.
The lowest dimensional example at $\Delta_0=6$ was
discussed in \secref{ssec:Dim234}.
It is annihilated by the extra symmetry generator $U_2$, i.e. it is 
unpaired.
Note, that for all of these operators two of the impurities 
are always next to each other. 

The operators $\Op$ are not exact planar eigenoperators of $D_4$. 
Nevertheless, we can project $D_4\Op$ to the piece 
proportional to $\Op$ and find $-36N^2$ as the coefficient. 
Consequently we have found a sequence 
of operator with planar dimensions
\[\label{eq:special3dim}
\Delta=\Delta_0+\frac{3\gym^2 N}{4\pi^2}-\frac{9\gym^4 N^2}{64\pi^4}.
\]
Interestingly, the anomalous dimension does not depend on the
bare dimension $\Delta_0$. In that sense, this operator behaves much like 
the highest mode in the series of 
two impurity operators discussed in 
\secref{sec:twoimp}. Assuming the
highest mode number were $n=(J+1)/2$ 
\footnote{The corresponding operator does not exist.
The highest mode number is $[J/2]$, for which, however,
the anomalous dimension does not differ considerably.},
the anomalous dimension \eqref{andim}
would equal $\gym^2N/\pi^2-\gym^4N^2/4\pi^4$,
which is also independent of $\Delta_0$. 
Furthermore, the frequency of the phase factor 
in \eqref{OfiniteJ} would also be extremal as in \eqref{eq:special3op}.

For a few of the lower dimensional operators we can also evaluate 
the anomalous dimensions up to four-loops 
(subject to the assumptions in deriving the 
corresponding vertices). Here, the yet undetermined coefficient
$\alpha$ does not contribute.
Intriguingly we find 
\[
\Delta=\Delta_0 + \frac{3\gym^2N}{4\pi^2} - \frac{9\gym^4N^2}{64\pi^2} + \frac{66\gym^6N^3}{1024\pi^6} 
- \frac{645\gym^8N^4}{16384\pi^8},
\]
for $\Delta_0=10,12,14$.
For $\Delta_0=8$ the four-loop contribution differs
\[
\Delta=8 + \frac{3\gym^2N}{4\pi^2} - \frac{9\gym^4N^2}{64\pi^2} + \frac{66\gym^6N^3}{1024\pi^6} 
- \frac{648\gym^8N^4}{16384\pi^8},
\]
while for $\Delta_0=6$ also the three-loop contribution is modified
\[
\Delta=6 + \frac{3\gym^2N}{4\pi^2} - \frac{9\gym^4N^2}{64\pi^2} + \frac{63\gym^6N^3}{1024\pi^6} 
- \frac{621\gym^8N^4}{16384\pi^8}.
\]
So the following picture emerges: 
The $k$-loop anomalous dimensions for the operators
of dimensions $\Delta_0=2+2k+n$, $n\geq 0$
seem to be equal.

\section{Conclusions and outlook}
\label{sec:results}

The main message of this paper is the proposal that perturbative
scaling dimensions in 4D conformal gauge theories should \emph{not} be 
computed on a case-by-case basis, using the standard, very laborious 
procedure. The latter consists in first working out classical and quantum 
two-point functions of a set of fields, subsequently
recursively diagonalizing and renormalizing the fields at each
order in the gauge coupling $\gym$ in order to find ``good'' conformal
fields satisfying the expected diagonal form in eq.\eqref{two} to
the desired order, and finally extracting the scaling dimensions from
eq.\eqref{two}. Instead one should focus on the dilatation operator
relevant to the general (i.e.~possessing an arbitrary engineering 
dimension) class of operators under study. Once it is found,
to the desired order in $\gym$,
the computation of the dilatation matrix becomes a straightforward, purely
\emph{algebraic} exercise. The subsequent calculation of the eigenvalues
of the matrix then yields the scaling dimensions, while the
eigenvectors resolve the mixing problem. 

In the present work we
illustrated this proposal in a specific example: We obtained the
dilatation operator up to ${\cal O}(\gym^4)$ of ${\cal N}=4$ $\grSU(N)$
Yang-Mills theory for arbitrary traceless pure scalar
fields, \emph{cf}.\ eqs.\eqref{introD4}. However, we certainly believe
that our methodology is rather general and should therefore be equally
applicable to other four-dimensional conformal gauge theories with
less supersymmetry.

In order to save us a significant amount of additional work we did use 
non-\-re\-normalization theorems as well as recent results
on the scaling dimensions of the so-called BMN two-impurity operators
\cite{Berenstein:2002jq,Gross:2002su,Santambrogio:2002sb} to fix
constants in our two-loop dilatation operator. Incidentally, this
may serve as an interesting illustration how the AdS/CFT 
correspondence (here in its latest reincarnation, i.e.~the
plane wave/BMN correspondence) can lead to new insights into
gauge theory. We then went on
to apply our example to a large number of situations of interest,
obtaining with ease a host of novel specific results 
for anomalous dimensions and for the resolution of mixing of 
scalar operators.
This yields important new information on the generic structure
of ${\cal N}=4$: E.g.~the degeneracy of certain single and double-trace
operators leads to unexpected $\frac{1}{N}$ terms in the associated
anomalous dimensions, and to the breakdown of a well-defined double expansion
in the 't Hooft coupling $\lambda=\gym^2 N$ and the genus counting
parameter $\frac{1}{N^2}$. One surely very
interesting issue we did not yet address is the interpretation of
these effects in the light of the AdS/CFT correspondence.

Our new results on $\superN=4$ SYM  already lead to new insights into the
plane wave strings/BMN correspondence. We were able to extend our quantum
mechanical description of BMN gauge theory~\cite{Beisert:2002ff},
deriving
from the two-loop dilatation operator the two-loop contribution to the
quantum Hamiltonian. This contribution strongly hinted at an all 
genera version of the celebrated all loop BMN square root formula, 
\emph{cf}.\ eqn.~\eqref{eq:BMNallloop}. Whether 
this generalization withstands closer scrutiny remains to be seen, of course.

It would also be very helpful to develop techniques similar to the
ones presented here for the efficient evaluation of correlation
functions of more than two local fields. It is e.g.~well known
that conformal invariance completely fixes the space-time form
of three-point functions ($x_{ij}=x_i-x_j$)
\[
\bigvev{\Op_\alpha(x_1) \Op_\beta(x_2) \Op_\gamma(x_3) }
=\frac{C_{\alpha \beta \gamma}}
{| x_{12}|^{\Delta_\alpha+\Delta_\beta-\Delta_\gamma}
| x_{23}|^{\Delta_\beta+\Delta_\gamma-\Delta_\alpha}
| x_{31}|^{\Delta_\gamma+\Delta_\alpha-\Delta_\beta}}.
\label{three}
\]
Therefore the quantities of interest are the finite structure
constants $C_{\alpha \beta \gamma}$ alone.
These constants appear in the operator product expansion
and apart from the scaling dimensions they
are the other central quantity in a conformal field theory of
local operators%
\footnote{
A number of structure constants
for two and three impurity BMN operators
\cite{Constable:2002hw,Chu:2002pd,Beisert:2002bb,
Constable:2002vq,Georgiou:2003aa}
have been obtained so far. 
In terms of the BMN correspondence it
would be important to understand their
dual on the string theory side \cite{Chu:2003qd}.}. 
However, in order to really obtain this form one has to use the correct
eigenstates of the dilatation operator. Furthermore, standard perturbative
computations are ``contaminated'' by useless finite and divergent 
contributions from the perturbative expansion of the 
weights $\Delta_\alpha$. One therefore wonders whether one may
generalize our methodology and develop purely algebraic techniques 
for directly finding the structure constants.
Of equal interest are four-point functions, which also seem
to possess many unexpected, simplifying features waiting to be
explored (see e.g.~\cite{Beisert:2002bb,Arutyunov:2003ae}).

By exploiting the recently discovered description of the planar
limit of the one-loop dilatation operator as the Hamiltonian of an 
integrable spin chain~\cite{Minahan:2002ve} 
we pointed out the existence of a new axial symmetry
of ${\cal N}=4$ Super Yang-Mills theory at $N=\infty$, linking fields 
of opposite parity (w.r.t.~reversing the order of fields inside
SU$(N)$ traces). 
We were able to prove the symmetry as a direct consequence of
the presence of non-trivial charges commuting with the Hamiltonian.
We furthermore derived this to follow immediately from
the existence of an $R$-matrix
satisfying the Yang-Baxter equation,
i.e.~from the integrability discovered in \cite{Minahan:2002ve}. 
An exciting open question is 
the interpretation of this symmetry (and of the integrability in
general) from the point of view of the planar gauge theory, or possibly from
the point of view of its dual string description.
This is even more pressing as we found much evidence that this integrable
structure extends to two loops (we did prove that at least two further
charges commute with the two loop dilatation operator) 
and it is obviously tempting to
conjecture that it holds to all loops.
One could therefore hope that a proper understanding of the integrability
might lead to the exact construction of the all-order planar
dilatation generator, see also the discussion below.

It would be important to extend the methods of this work
to fix the remaining terms 
in the dilatation operator eq.\eqref{dilexp} pertaining to further 
classes of operators such as scalars with $\grSO(6)$ traces, 
fermions, field strengths or covariant derivatives.
The latter would be interesting in order to study the 
high spin limit of \cite{Gubser:2002tv}, see also 
\cite{Kotikov:2003fb}. 
Here we expect the superconformal symmetry of the model to be helpful.
It would also be fascinating to investigate whether the 
reformulation of the planar theory 
as an integrable spin chain can be extended to include the other fields.

It is striking that, once the dilatation operator is found, the 
calculations of anomalous dimension matrices become purely
algebraic. However, in order to fully justify its derivation
we did need to take a look at two-point functions 
(\emph{cf}.\ \secref{sec:oneloop},\ref{sec:twoloop}). It is natural to wonder whether
there are further shortcuts that leads to the determination of the
terms in the perturbative expansion of the dilatation operator 
eq.\eqref{dilexp}. For the ${\cal N}=4$ model this is not an
unreasonable expectation: First of all, the theory is scale-invariant
on the quantum level and therefore possesses a finite dilatation
operator. Thus one
could expect that the renormalization procedure of 
\secref{sec:oneloop},\ref{sec:twoloop} is only a scaffold that
one might be able to avoid. Secondly, the theory's action is unique,
and entirely determined by the maximal superconformal symmetry
$\grSU(2,2|4)$. Is there a way to use the symmetry algebra,
possibly paired with some further insights into the ${\cal N}=4$ model,
to completely fix the structure of the dilatation operator \eqref{dilexp}? 

We would certainly like to push our procedure to higher loops.
This is not a 
pointless exercise. E.g.~little seems to be known about
the analytic structure of the \emph{exact} anomalous dimension
of low dimensional fields such as Konishi. All we currently have is
the two-loop result eq.\eqref{Kontwoloop}, the planar
three-loop conjecture~eq.\eqref{Konthreeloop}
 and the conjectured 
(from AdS/CFT) strong coupling behavior 
$\Delta \sim (\gym^2 N)^{1/4}$. Knowing further terms
in the perturbative expansion might give essential clues about
the convergence structure of the series (is the radius of
convergence zero or finite?) and might allow us to estimate
the strong coupling result by Pad\'e approximants. Incidentally,
the Konishi field eq.\eqref{dimtwo} is particularly interesting
as it cannot mix with any other fields and is therefore known to
be an exact eigenstate of the dilatation operator to all
orders in perturbation theory.

We have taken  first steps towards constraining the complete dilatation
operator, building directly on 
the consequences of
the spin chain picture: Assuming the above mentioned axial 
symmetry to
hold also beyond two loops we were able to derive a planar version of
the three-loop dilatation operator which in turn allowed us to obtain
further results on anomalous dimensions. A preliminary study of the
possible planar four-loop structures showed that the information which
completely fixed the three-loop case leaves one coefficient of the
planar four-loop dilatation operator undetermined. A speculation of,
potentially, tremendous importance would be that imposing the full
integrability structure completely fixes, at each loop order, the exact
planar dilatation operator.

\subsection*{Acknowledgments}

We would like to thank Natan Andrei, Gleb Arutyunov, David Berenstein, 
Volodya Kazakov, Thomas Klose, Stefano Kovacs, 
Ari Pankiewicz, Jan Plefka and Thomas Quella for interesting
discussions. M.S. thanks the Rutgers Physics department for hospitality while
working on parts of this manuscript.
N.B.~dankt der \emph{Studienstiftung des
deutschen Volkes} f\"ur die Unterst\"utzung durch ein
Promotions\-f\"orderungsstipendium.

\appendix

\section{Conventions}
\label{sec:conv}

We use the following $\superN=4$ supersymmetric action in components
\<\label{eq:SYMaction}
S\eq\frac{1}{2}\int\frac{d^{4-2\epsilon}x}{(2\pi)^{2-\epsilon}}
\Tr \Big(
\quarter F_{\mu\nu}F_{\mu\nu}
+\half \cder_\mu\Phi_m \cder_\mu\Phi_m
-\quarter\gnorm^2\mu^{2\epsilon}\,\comm{\Phi_m}{\Phi_n}\comm{\Phi_m}{\Phi_n} 
\nl\qquad\qquad\qquad\qquad\qquad
+\half\Psi^\trans\Sigma_\mu D_\mu\Psi
-\sfrac{i}{2} \gnorm\mu^{\epsilon}\,\Psi^\trans \Sigma_m \comm{\Phi_m}{\Psi}
\Big),
\nln
\cder_\mu X\eq\partial_\mu X-i\gnorm\mu^{\epsilon} \comm{A_\mu}{X},
\nln
F_{\mu\nu}\eq\partial_\mu A_\nu-\partial_\nu A_\mu-i\gnorm\mu^{\epsilon} \comm{A_\mu}{A_\nu}.
\>
The coupling constant $\gnorm$ is related to the common coupling 
constant of $\superN=4$ SYM by 
\[\label{eq:gnorm}
\gnorm^2=\frac{\gym^2}{4(2\pi)^{2-\epsilon}}\to \frac{\gym^2}{16\pi^2}.\]
This normalization turns out convenient when evaluating
space-time integrals and when operators with fermions and derivative insertions 
are considered. 

The fields carry a color structure, $\Phi_m=T^a \Phi_m^{(a)}$.
We will consider the gauge group $\grU(N)$ whose generators $T^a$
we have normalized such that 
\[\Tr T^a T^b=\delta^{ab}, \qquad
\sum\nolimits_{a} (T^a)^{\alpha}_{\,\,\,\beta} (T^a)^{\gamma}_{\,\,\,\delta}=
\delta_{\delta}^{\alpha}\delta_{\beta}^{\gamma}.  \]
This implies the $\grU(N)$ fusion and fission rules
\[
\Tr T^a A \Tr T^a B=\Tr AB,\quad
\Tr T^a A T^a B=\Tr A\Tr B.
\]

In this work we make extensive use of variations
with respect to a field $\Phi_m$, which we will denote by
\[\check \Phi_m=
\frac{\delta}{\delta \Phi_m}=
T^a\frac{\delta}{\delta \Phi_m^{(a)}}.
\]
It is therefore understood when the variation hits a field, both the
variation symbol and the field are replaced by a color generator $T^a$.
Note that the variation symbols $\check \Phi_m$ act only to the right.
In a normal ordered word of 
fields and variations,
$\normord{\Phi\check\Phi\Phi\Phi\ldots\,}$,
it is understood that the variations do not contract to
any of the fields within the normal ordering. 

\section{Scalar space-time integrals}
\label{sec:integrals}

We have normalized the scalar propagator to
\[\label{eq:prop}
I_{xy}=\frac{\Gamma(1-\epsilon)}{\bigabs{\half (x-y)^2}^{1-\epsilon}},\]
it is a rather convenient normalization when derivatives are taken
(e.g. for fermions). The following integrals are required at one-loop
\<\label{eq:YXH}
Y_{x_1x_2x_3}\eq\mu^{2\epsilon}\int\frac{d^{4-2\epsilon}z}{(2\pi)^{2-\epsilon}} I_{x_1z}I_{x_2z}I_{x_3z},
\nln
X_{x_1x_2x_3x_4}\eq\mu^{2\epsilon}\int\frac{d^{4-2\epsilon}z}{(2\pi)^{2-\epsilon}} I_{x_1z}I_{x_2z}I_{x_3z}I_{x_4z},
\\\nn
\tilde H_{x_1x_2,x_3x_4}\eq\half\mu^{2\epsilon}\lrbrk{\frac{\partial}{\partial x_1}+\frac{\partial}{\partial x_3}}^2
\int\frac{d^{4-2\epsilon}z_1\,d^{4-2\epsilon}z_2}{(2\pi)^{4-2\epsilon}} 
I_{x_1z_1}I_{x_2z_1}I_{z_1z_2}I_{z_2x_3}I_{z_2x_4}.
\>
When evaluated in two-point functions they yield \cite{Kazakov:1984ns}
\<\label{eq:YXH2pt}
\frac{Y_{00x}}{I_{0x}}\eq
\frac{1}{\epsilon(1-2\epsilon)}\,
\xi,
\nln
\frac{X_{00xx}}{I_{0x}^2}\eq
\frac{2(1-3\epsilon)\gamma}{\epsilon(1-2\epsilon)^2}\,\xi,
\\\nn
\frac{\tilde H_{0x,0x}}{I_{0x}^2}\eq
-\frac{2(1-3\epsilon)(\gamma-1)}{\epsilon^2(1-2\epsilon)}
\,\xi,
\>
where
\[\label{eq:xigamma}
\xi=\frac{\Gamma(1-\epsilon)}{\bigabs{\half\mu^2 x^2}^{-\epsilon}},
\qquad
\gamma=\frac{\Gamma(1-\epsilon)\Gamma(1+\epsilon)^2\Gamma(1-3\epsilon)}
{\Gamma(1-2\epsilon)^2\Gamma(1+2\epsilon)}=1+6\zeta(3)\epsilon^3+\order{\epsilon^4}.
\]

\section{Renormalization at higher loops}
\label{sec:renorm}

To obtain the arbitrary loop correlator we insert
all $l$-loop connected Green functions $W_{2l}$ 
in the correlator
\[\label{eq:allloopW}
\bigvev{\Op_\alpha^+ \Op_\beta^-}=
\bigeval{\exp(W_0)
\exp\lrbrk{\tsum_{l=1}^\infty \gnorm^{2l} W_{2l}(x,\check\Phi^+,\check\Phi^-)}\Op_\alpha^+\Op_\beta^- }_{\Phi=0}.
\]
In analogy to \eqref{eq:oneloopV} we change 
the argument $\check \Phi^+$ of $W_{2l}(x,\check\Phi^+,\check\Phi^-)$
to $I^{-1}_{0x}\Phi^-$
\[\label{eq:allloopWsub}
\bigvev{\Op_\alpha^+ \Op_\beta^-}=
\bigeval{\exp(W_0)
\normord{\exp\lrbrk{\tsum_{l=1}^\infty \gnorm^{2l} W_{2l}(x,I^{-1}_{0x}\Phi^-,\check\Phi^-)}}
\Op_\alpha^+\Op_\beta^- }_{\Phi=0}.
\]
Alternatively, we could change the argument
$\check \Phi^-$ to $I^{-1}_{0x}\Phi^+$.
This does not make a difference as
$W_{2l}(x,\check\Phi^+,\check\Phi^-)$
is symmetric in the arguments $\check\Phi^+$ and $\check\Phi^-$.
We would then like to rewrite
\eqref{eq:allloopWsub} in a convenient form for
the conformal structure of the correlator:
\[\label{eq:nloopV}
\bigvev{\Op_\alpha^+ \Op_\beta^-}=
\bigeval{\exp(W_0)
\exp\bigbrk{V^-(x)}\Op_\alpha^+ \Op_\beta^- }_{\Phi=0},
\]
where 
\[\label{eq:Vndef}
V(x)=\sum_{l=1}^\infty \gnorm^{2l}V_{2l}(x) -\sfrac{1}{48}\gnorm^8\bigcomm{V_2(x)}{\comm{V_2(x)}{V_4(x)}}+\ldots.
\]
The terms $V_{2l}$ are defined by the equality of 
\eqref{eq:allloopWsub} and \eqref{eq:nloopV}
\[\label{eq:Vn}
\exp\bigbrk{V(x)}=
\normord{\exp\lrbrk{\tsum_{l=1}^\infty \gnorm^{2l} W_{2l}(x,I^{-1}_{0x}\Phi,\check\Phi)}},
\]
which will have to be solved perturbatively. All
the terms that arise due to normal ordering of the exponential
and the commutator terms in \eqref{eq:Vndef} need to
be absorbed into the definition of higher order vertices.
For example, the two-loop effective vertex is
\[\label{eq:V4}
V_4(x)=\normord{W_4(x,I_{0x}^{-1}\Phi,\check\Phi)}
-\half \bigbrk{V_2(x)V_2(x)-\normord{V_2(x)V_2(x)}}
.\]
The commutator terms in \eqref{eq:Vndef} were included for 
convenience, we will explain this issue below.
The symmetry of $W_{2l}$ is translated into the 
effective equality
\[\label{eq:Vsym}
\exp(W_0)V^-_{2l}(x)=\exp(W_0)V^+_{2l}(x).
\]
In fact we can introduce a transpose operation 
on a generator $X$ by the definition 
\[\label{eq:transpose}
\exp(W_0)\,X^-=\exp(W_0)\,X^{\trans\, +},
\]
In other words, letting $X$ act on $\Phi^-$
is equivalent to letting $X^\trans$ act on $\Phi^+$.
The symmetry of the vertices translates to 
\[\label{eq:Vsymtranspose}
V^\trans_{2l}(x)=V_{2l}(x).
\]

We renormalize the operators according to 
\[\label{eq:Zn}
\tilde\Op=\exp\bigbrk{-\half Z(x_0)}\Op,
\]
with 
\[\label{eq:Zndef}
Z(x_0)=\sum_{l=1}^\infty \gnorm^{2l} V_{2l}(x_0)-\sfrac{1}{12}\gnorm^6 \bigcomm{V_2(x_0)}{V_4(x_0)}+\ldots
\]
This gives
\[\label{eq:nloopVrenorm}
\bigvev{\Op_\alpha^+ \Op_\beta^-}=
\bigeval{\exp(W_0)
\exp\bigbrk{V^-(x)}\exp\bigbrk{-\half Z^+(x_0)}\Op_\alpha^+ \exp\bigbrk{-\half Z^-(x_0)}\Op_\beta^- }_{\Phi=0}.
\]
We can commute objects with a $+$ and a $-$ index freely and 
use the transpose operation \eqref{eq:transpose}
to make $Z^+$ act on $\Phi^-$ instead.
We get
\[\label{eq:nloopVrenorm2}
\bigvev{\Op_\alpha^+ \Op_\beta^-}=
\bigeval{\exp(W_0)
\exp\bigbrk{-\half Z^{\trans-}(x_0)}\exp\bigbrk{V^-(x)}\exp\bigbrk{-\half Z^-(x_0)}\Op_\alpha^+ \Op_\beta^- }_{\Phi=0}.
\]
The vertices $V_{2l}(x_0)$ in $Z(x_0)$ are symmetric, \eqref{eq:Vsymtranspose}, 
only the commutator in \eqref{eq:Zndef} requires special 
care, because $V_2$ and $V_4$ need to be transformed
consecutively. This effectively inverts their order and
flips the sign of the commutator:
\[\label{eq:Ztndef}
Z^\trans(x_0)=\sum_{l=1}^\infty \gnorm^{2l} V_{2l}(x_0)+\sfrac{1}{12}\gnorm^6 \bigcomm{V_2(x_0)}{V_4(x_0)}+\ldots
\]
In a renormalizable field theory the dependence of $V_{2l}$ on $x$ is
determined, we write
\[\label{eq:Vnx}
V_{2l}(x)=\xi^l V_{2l},\qquad
\xi=\frac{\Gamma(1-\epsilon)}{\bigabs{\half\mu^2 x^2}^{-\epsilon}}.
\]
We combine the last three exponentials in \eqref{eq:nloopVrenorm2}
into one with exponent
\[\label{eq:nloopVrenormexp}
\sum_{l=1}^{\infty} (\xi^l-\xi_0^l) \gnorm^{2l}V^-_{2l} 
-\sfrac{1}{48}\gnorm^8(\xi-\xi_0)^4\bigcomm{V_2^-}{\comm{V_2^-}{V_4^-}}
+\ldots
\]
The $l$-loop Green function $W_{2l}$ is expected to have multiple poles
at $\epsilon=0$.
In a conformal field theory, however, these poles
must have cancelled in the combination $V_{2l}$
as given by \eqref{eq:Vndef}, \eqref{eq:Vn}.
If so, we can finally send the regulator to zero and find 
\[\label{eq:nloopDrenorm}
\bigvev{\tilde\Op_\alpha^+ \tilde\Op_\beta^-}=
\bigeval{\exp(W_0)
\exp\bigbrk{\log (x_0^2/x^2)\tsum_{l=1}^\infty\gnorm^{2l} D_{2l}^-}\Op_\alpha^+ \Op_\beta^- }_{\Phi=0},
\]
with 
\[\label{eq:Dn}
D_{2l}=-l\lim_{\epsilon\to 0} \epsilon V_{2l}.
\]
Note that the commutator term in 
\eqref{eq:nloopVrenormexp} vanishes due to four powers
of $\epsilon$ from $(\xi-\xi_0)^4$ opposed to only three powers
of $1/\epsilon$ from the $V_{2l}$.
For this cancellation to happen the commutator terms in 
\eqref{eq:Vndef} and \eqref{eq:Zndef} are necessary%
\footnote{We have investigated all possible terms that arise in a four-loop
computation. We found that
exactly the commutator structure in \eqref{eq:Vndef} 
was required to obtain a finite, conformally covariant
correlator.}.

Some comments about the renormalization program are
in order. 
The effective vertices $V_{2l}$ are connected diagrams.
They are generated from the Green functions $W_{2l}$ 
by removing the normal ordering of an exponential
\eqref{eq:Vn}
and adding commutators \eqref{eq:Vndef}.
One can easily convince oneself that these operations
produce connected diagrams.
The same is then true also for the dilatation generator $D$.
Secondly, the program ensures that the coefficient
of the two-point function is given by 
free-contractions of the unrenormalized operators.
Thirdly, the effective vertices $V_{2l}$ are 
symmetric with respect to the scalar product induced
by free contractions, see \eqref{eq:Vsym}.
The same holds for the dilatation generator
which consequently has real eigenvalues%
\footnote{Some eigenvalues may appear to be complex.
This can only happen if the corresponding 
eigenvector has zero norm. In this case
the operator in fact does not exist.
This happens if the rank of the group is small
compared to the size of the operator. 
Then group identities make some operators
linearly dependent.}.

\section{A collection of anomalous dimensions}
\label{sec:anodim}

\subsection{Away from the unitarity bounds}
\label{ssec:OffUnitarity}

The following operators have a form similar to
the ones discussed in \secref{sec:intro}.
As they contain all six scalar fields and $\grSO(6)$
traces they mix with operators containing fermions and
derivatives,
see \secref{ssec:purescalars}.
This mixing, however, only becomes relevant 
beyond one-loop and we can determine 
their one-loop anomalous dimensions using 
\eqref{eq:D2}.

\subsubsection{Dimension 5, $[0,1,0]$}

We find one single trace operator with dimension 
\setcounter{equation}{-1}
\[\Delta=5+\frac{5\gym^2N}{4\pi^2},\]
three single-trace and three double-trace operators
with dimensions 
\[\Delta=5+\frac{\gym^2N}{4\pi^2}\,\omega,\]
where $\omega$ is a root of the sixth-order equation
\<
&&\omega^6
-17\omega^5 
+\lrbrk{110-\frac{50}{N^2}}\omega^4 
-\lrbrk{335-\frac{565}{N^2}}\omega^3 
+\lrbrk{475-\frac{2440}{N^2}+\frac{400}{N^4}}\omega^2 
\nn\\&&\phantom{\omega^6}
-\lrbrk{250-\frac{4850}{N^2}+\frac{1600}{N^4}}\omega  
-\lrbrk{\frac{3750}{N^2}-\frac{4000}{N^4}}=0.
\>
Two of these operators,
a single-trace and a double-trace one, 
are degenerate in the planar limit.
The degeneracy is lifted by $\frac{1}{N}$ corrections
and the t' Hooft expansion of their 
scaling dimension is subject to the
issues discussed in \secref{ssec:degen}.

\subsubsection{Dimension 6, $[0,2,0]$, planar}

We find one pair of operators with
dimension 
\[\Delta=6+\frac{7\gym^2N}{8\pi^2},\]
two operators with dimension 
\[\Delta=6+\frac{\gym^2N}{\pi^2},\]
and six operators with dimensions
\[\Delta=6+\frac{\gym^2N}{8\pi^2}\,\omega,\]
where $\omega$ is a root of the sixth-order equation
\[\omega^6-43\omega^5+731\omega^4-6238\omega^3
+27936\omega^2-61776\omega+52272 =0.\]

\subsubsection{Dimension 6, $[1,0,1]$, planar}

We find two pairs of operators with
dimension 
\[\Delta=6+\frac{\gym^2N}{\pi^2},\]
two operators with dimension 
\[\Delta=6+\frac{5\gym^2N}{4\pi^2},\]
and three operators with dimensions
\[\Delta=6+\frac{\gym^2N}{8\pi^2}\,\omega,\]
where $\omega$ is a root of the cubic equation
\[\omega^3 -23\omega^2 +158\omega  -308 =0.\]

\subsection{Further away from the unitarity bounds}
\label{ssec:OffUnitarity2}

\subsubsection{Dimension 6, $[0,0,0]$, planar}

We find five operators with dimensions
\[\Delta=6+\frac{\gym^2N}{8\pi^2}\,\omega,\]
where $\omega$ is a root of the quintic equation
\[\omega^5-43\omega^4 +701\omega^3 
-5338\omega^2 +18480\omega -21960 =0.\]

\subsection{Three impurities}

The following operators are in the $\grSO(6)$ 
representations for which mixing with fermions
and derivative insertions is prohibited,
see \secref{ssec:purescalars}.
We may apply the dilatation generator
at two-loops \eqref{eq:D4}.

\subsubsection{Dimension 8, $[3,2,3]$, planar}
We find a pair of operators with dimension
\[\label{eq:dim8pair}
\Delta=8+\frac{\gym^2N}{2\pi^2}-\frac{5\gym^4 N^2}{64\pi^4},
\]
and a single operator with dimension
(see \secref{ssec:3tower})
\[
\Delta=8+\frac{3\gym^2N}{4\pi^2}-\frac{9\gym^4 N^2}{64\pi^4}.
\]

\subsubsection{Dimension 9, $[3,3,3]$, planar}
We find three pairs of operators with 
dimensions
\[\Delta=9+\frac{\gym^2 N}{16\pi^2}\,\omega,\]
where $\omega$ is a root of the cubic equation
\[
\omega^3-34\omega^2+360\omega-1176+\frac{\gym^2N}{16\pi^2}(102\omega^2-2100\omega+9912)=0.
\]

\subsubsection{Dimension 10, $[3,4,3]$, planar}
We find one operator with dimension
(see \secref{ssec:3tower})
\[
\Delta=10+\frac{3\gym^2 N}{4\pi^2}-\frac{9\gym^4N^2}{64\pi^4},
\]
and three pairs of operators with
dimensions
\[\Delta=10+\frac{\gym^2 N}{16\pi^2}\,\omega,\]
where $\omega$ is a root of the cubic equation
\[
\omega^3-30\omega^2+276\omega-768+\gym^2N(86\omega^2-1516\omega+5984)=0.
\]

\subsection{Four impurities}
\subsubsection{Dimension 8, $[4,0,4]$, planar}

We find three operators with dimensions
\[\Delta=8+\frac{\gym^2 N}{16\pi^2}\,\omega,\]
where $\omega$ is a root of the cubic equation
\[
\omega^3-40\omega^2+464\omega-1600+\frac{\gym^2 N}{16\pi^2}(128\omega^2-2720\omega+12800)=0.
\]

\subsubsection{Dimension 9, $[4,1,4]$, planar}

We find a pair of operators with dimension%
\footnote{This dimension matches \eqref{eq:pairdim}.}
\[
\Delta=9+\frac{5\gym^2N}{8\pi^2}-\frac{15\gym^4N^2}{128\pi^4},
\]
and two operators with dimensions
\[\Delta=9+\frac{\gym^2N\lrbrk{3\pm\sqrt{3}}}{4\pi^2}
-\frac{\gym^4N^2\lrbrk{18\pm 9\sqrt{3}}}{128\pi^4}\]

\subsubsection{Dimension 10, $[4,2,4]$, planar}

We find two pairs of operators with dimension
\[
\Delta=10+\frac{\gym^2N\lrbrk{11\pm\sqrt{5}}}{16\pi^2}
-\frac{\gym^4N^2\lrbrk{31\pm3\sqrt{5}}}{256\pi^4},
\]
and six operators with dimensions
\[\Delta=10+\frac{\gym^2N}{4\pi^2}\,\omega,\]
where $\omega$ is a root of the sixth order equation
\<
&&\omega^6-21\omega^5+173\omega^4-711\omega^3+1525\omega^2-1603\omega+637
\\\nn&&
\phantom{\omega^6}+\frac{\gym^2N}{16\pi^2}
\lrbrk{67\omega^5-1074\omega^4+6409\omega^3-17623\omega^2+22078\omega-9947}=0.
\>

\subsection{Five Impurities}

\subsubsection{Dimension 10, $[5,0,5]$, planar}

We find four operators with dimensions
\[\Delta=10+\frac{\gym^2N}{4\pi^2}\,\omega,\]
where $\omega$ is a root of the quartic equation
\<
&&\omega^4-15\omega^3+78\omega^2-165\omega+120
\\\nn&&
\phantom{\omega^6}+\frac{\gym^2N}{16\pi^2}
\lrbrk{47\omega^3-472\omega^2+1430\omega-1305}=0.
\>
%

\section {The operator $\delta D_4$ \label{deltaD4}}
\label{sec:deltaD4}

Here we list the exact contributions to the generator $\delta D_4$
acting on $\Op_p^{J_0;J_1,\ldots,J_k}$,
\emph{cf}.\ equations~\eqref{deltaD} and~\eqref{D4}. 
We see that in strong contrast to $(D_2)^2$, the operator $\delta D_4$
only creates states in the combination $\Op_1-\Op_0$; we
write in short
\[
\Op^{J_0;J_1,\ldots,J_k}_{1-0}=
\Op^{J_0;J_1,\ldots,J_k}_{1}-\Op^{J_0;J_1,\ldots,J_k}_{0}.
\]
The operators with $J_0=0$ are always annihilated, 
we assume $J_0>1$. The planar part is 
\[
\delta {D}_{4;0}\, \Op_p^{J_0;J_1,\ldots,J_k}=
4(\delta_{p,0}+\delta_{p,J_0}-\delta_{p,1}-\delta_{p,J_0-1})\,
\Op^{J_0;J_1,\ldots,J_k}_{1-0}.
\]
The operator can split off one trace
\<\label{D4+}
\delta {D}_{4;+}\, \Op_p^{J_0;J_1,\ldots,J_k}\eq
4\delta_{p\neq 0,p\neq J_0}
\bigbrk{\Op^{p;J_1,\ldots,J_k,J_0-p}_{1-0}+\Op^{J_0-p;J_1,\ldots,J_k,p}_{1-0}}
\nl
-8\delta_{p>1}\Op^{J_0-p+1;J_1,\ldots,J_k,p-1}_{1-0}
\nl
-8\delta_{p<J_0-1}\Op^{p+1;J_1,\ldots,J_k,J_0-p-1}_{1-0}
\nl
+4(\delta_{p,0}+\delta_{p,J_0})\sum_{J_{k+1}=1}^{J_0-1}
\Op^{J_0-J_{k+1};J_1,\ldots,J_{k+1}}_{1-0},
\>
or two traces 
\<
\label{D4++}
{\delta D_{4;++}}\, \Op_p^{J_0;J_1,\ldots,J_k}\eq
4\delta_{p\neq 0}\sum_{J_{k+1}=1}^{J_0-p-1}
\Op^{J_0-p-J_{k+1};J_1,\ldots,J_{k+1},p}_{1-0}
\nl
+4\delta_{p\neq J_0}\sum_{J_{k+1}=1}^{p-1}
\Op^{p-J_{k+1};J_1,\ldots,J_{k+1},J_0-p}_{1-0} 
\nl
-4\sum_{J_{k+1}=1}^{J_0-p-2}
\Op^{p+1;J_1,\ldots,J_{k+1},J_0-p-J_{k+1}-1}_{1-0}
\nl
-4\sum_{J_{k+1}=1}^{p-2} 
\Op^{J_0-p+1;J_1,\ldots,J_{k+1},p-J_{k+1}-1}_{1-0},
\>
it can join two traces
\[\label{D4-} 
\delta {D}_{4;-}\, \Op_p^{J_0;J_1,\ldots,J_k}= 
4(\delta_{p,0}+\delta_{p,J_0}) \sum_{i=1}^k J_i\, 
\Op^{J_0+J_i;J_1,\ldots,\makebox[0pt]{\,\,\,\,$\times$}J_{i},\ldots,J_k}_{1-0},
\]
or it can let two traces interact without
changing the number of traces
\<\label{D400}
\delta {D}_{4;+-}\, \Op_p^{J_0;J_1,\ldots,J_k}\eq
4\delta_{p\neq 0}
\sum_{i=1}^k J_i\, 
\Op^{J_0+J_i-p;J_1,\ldots,\makebox[0pt]{\,\,\,\,$\times$}J_{i},\ldots,J_k,p}_{1-0}
\nl
+4\delta_{p\neq J_0}\sum_{i=1}^k J_i\, 
\Op^{J_i+p;J_1,\ldots,\makebox[0pt]{\,\,\,\,$\times$}J_{i},\ldots,J_k,J_0-p}_{1-0}
\nl
-4\sum_{i=1}^k J_i\, 
\Op^{p+1;J_1,\ldots,\makebox[0pt]{\,\,\,\,$\times$}J_{i},\ldots,J_k,J_0+J_i-p-1}_{1-0}
\nl
-4\sum_{i=1}^k J_i\, 
\Op^{J_0-p+1;J_1,\ldots,\makebox[0pt]{\,\,\,\,$\times$}J_{i},\ldots,J_k,J_i+p-1}_{1-0}
.
\>
%

\section{Charges of the spin chain}
\label{sec:highercharges}

In this appendix we present some of the commuting charges of
the non-nearest neighbor $\grSU(2)$ spin chain 
investigated in \secref{sec:highergrounds}. We use the notation 
\[\{n_1,n_2,\ldots\}=
\sum_{k=1}^L P_{k+n_1,k+n_1+1}
P_{k+n_2,k+n_2+1}\ldots\]
We make extensive use of the identity
\<\label{eq:U2indent}
\{\ldots,n,n\pm 1,n,\ldots\}\eq\{\ldots,\ldots\}
-\{\ldots,n,\ldots\}-\{\ldots,n\pm 1,\ldots\}
\nl
+\{\ldots,n,n\pm 1,\ldots\}+\{\ldots,n\pm 1,n,\ldots\},
\>
due the impossibility of antisymmetrizing three sites in $\grSU(2)$.
The following expression for the higher charges are unique
up lower charges multiplied by powers of the coupling constant.
\bigskip

\noindent\textbf{\mathversion{bold}The first charge $D=Q_1$}
\<
D_0\eq \{\},
\nln
D_2\eq 2\{\}-2\{0\},
\nln
D_4\eq -8\{\}+12\{0\}-2\lrbrk{\{0,1\}+\{1,0\}},
\nln
D_6\eq 
60\{\}
-104\{0\}
+24\lrbrk{\{0,1\}+\{1,0\}}
+4\{0,2\}
-4\lrbrk{\{0,1,2\}+\{2,1,0\}},
\nln
D_8\eq
+(-572 + 4\alpha )  \{\}
+(1072-12\alpha+4\beta)\{0\}
\nl
+(-278+4\alpha-4\beta) (\{0, 1\} + \{1, 0\})  
+(-84+6\alpha-2\beta)\{0, 2\}
-4 \{0, 3\}
\nl
+4(\{0, 1, 3\} + \{0, 2, 3\} + \{0, 3, 2\} + \{1, 0, 3\}) 
\nl
+(78+2\beta) (\{0, 1, 2\} + \{2, 1, 0\}) 
+(-6-4\alpha+2\beta) (\{0, 2, 1\} + \{1, 0, 2\})  
\nl
+(1-\beta) ( \{0, 1, 3, 2\} + \{0, 3, 2, 1\} + \{1, 0, 2, 3\} + \{2, 1, 0, 3\} )  
\nl
+(2\alpha-2\beta) \{1, 0, 2, 1\}
+2\beta (\{0, 2, 1, 3\} + \{1, 0, 3, 2\})  
\nl
-10(\{0, 1, 2, 3\} + \{3, 2, 1, 0\}).
\>

\noindent\textbf{\mathversion{bold}The second charge $U=Q_2$}
\<
U_2\eq 4\lrbrk{\{1,0\}-\{0,1\}},
\nln
U_4\eq 8\lrbrk{\{2,1,0\}-\{0,1,2\}},
\nln
U_6\eq
8\lrbrk{\{0, 1, 3\} + \{0, 2, 3\}-\{0, 3, 2\} - \{1, 0, 3\}}
\nl
+40 \lrbrk{\{0, 1, 2\}-\{2, 1, 0\}} 
+16\lrbrk{\{3, 2, 1, 0\}-\{0, 1, 2, 3\}}.
\>

\noindent\textbf{\mathversion{bold}The third charge $Q_3$}
\<Q_{3,2}\eq
- 2\{0\} 
+ \lrbrk{\{0,1\} + \{1,0\}} 
\nl
+ \lrbrk{\{0,2,1\} + \{1,0,2\}}
- \lrbrk{\{0,1,2\} + \{2,1,0\}},
\nln
Q_{3,4}\eq
-2\{0\}
+\lrbrk{\{0,1\}+\{1,0\}}
- 4\{0,2\} 
\nl
- 3\lrbrk{\{0,1,2\} + \{2,1,0\} }
+ 5\lrbrk{\{0,2,1\} + \{1,0,2\} }
+ 2\{1,0,2,1\} 
\nl
- 3\lrbrk{\{0,1,2,3\} + \{3,2,1,0\}}
+ \lrbrk{\{0,2,1,3\} + \{1,0,3,2\} }
\nl
+ \lrbrk{\{0,1,3,2\} + \{2,1,0,3\} + \{0,3,2,1\} + \{1,0,2,3\} }.
\>

\noindent\textbf{\mathversion{bold}The fourth charge $Q_4$}
\<Q_{4,2}\eq
-2\lrbrk{\{0, 1, 2\} - \{2, 1, 0\}} 
\nl
 -\lrbrk{\{0, 2, 1, 3\} - \{1, 0, 3, 2\}} 
 +\lrbrk{\{0, 1, 2, 3\} - \{3, 2, 1, 0\}}
\nl
 +\lrbrk{\{0, 3, 2, 1\} -\{0, 1, 3, 2\} - \{1, 0, 2, 3\} + \{2, 1, 0, 3\}}.
\>


\bibliography{dila}
\bibliographystyle{nb}

\end{document}